\documentclass[twocolumn]{aastex631}
\usepackage{bm}
\usepackage{amsmath}
\usepackage{hyperref}
\newcommand{\knrt}{\rm{yr^{-1}\ Mpc^{-3}}}
\newcommand{\knrtgiga}{\rm{yr^{-1}\ Gpc^{-3}}}

\newcommand{\knownbns}{GW170817}

\newcommand{\safKN}{SAF19}
\newcommand{\mosfit}{\texttt{MOSFiT}}
\newcommand{\redback}{{\sc Redback}}
\newcommand{\python}{{\texttt{python}}}
\newcommand{\msun}{{\rm{M_\odot}}}

\newcommand{\redbackdocs}{\url{https://redback.readthedocs.io/en/latest/}}
\newcommand{\opacitysym}{$\kappa$}
\newcommand{\opacityunit}{$\mathrm{cm}^{2}/\mathrm{g}$}

\newcommand{\mejone}{{M_{\mathrm{ej}1}}}
\newcommand{\mejtwo}{{M_{\mathrm{ej}2}}}
\newcommand{\mejthree}{{M_{\mathrm{ej}3}}}

\newcommand{\vejone}{{v_{\mathrm{ej}1}}}
\newcommand{\vejtwo}{{v_{\mathrm{ej}2}}}
\newcommand{\vejthree}{{v_{\mathrm{ej}3}}}

\newcommand{\kappaone}{{\kappa_{\mathrm{ej}1}}}
\newcommand{\kappatwo}{{\kappa_{\mathrm{ej}2}}}
\newcommand{\kappathree}{{\kappa_{\mathrm{ej}3}}}

\newcommand{\timedelay}{\Delta t}
\newcommand{\magnification}{\mu}
\newcommand{\delaytime}{\tau}

\newcommand{\pwrlw}{\alpha}

\newcommand{\srcred}{z_\mathrm{s}}
\newcommand{\lensred}{z_\mathrm{l}}
\newcommand{\sigmavd}{\sigma}
\newcommand{\einrad}{\theta_{\mathrm{Ein}}}
\newcommand{\maxsep}{\theta^{\mathrm{sep}}_{\mathrm{max}}}
\newcommand{\velunit}{\mathrm{km\, s^{-1}}}
\newcommand{\sigmalens}{\Omega_{\rm lens}}
\newcommand{\mutot}{\mu_{\rm tot}}

\newcommand{\hnot}{H_{0}}
\newcommand{\tpeak}{t_{\rm{p}}}
\newcommand{\tpeakthree}{t_{\rm{p+3}}}
\newcommand{\afterglow}{AT2017gfo}
\newcommand{\typeonea}{Type Ia}
\shorttitle{Lensed Kilonovae in LSST}
\shortauthors{Ganguly and More}
\graphicspath{{./}{figures/}}
\begin{document}
\title{Detectability of Gravitationally Lensed Kilonovae in the Rubin LSST}
\correspondingauthor{Anindya Ganguly}
\email{anindyaganguly12@gmail.com, anupreeta@iucaa.in}
\author[0000-0002-0991-8438]{Anindya Ganguly}
\affiliation{Inter-University Centre for Astronomy and Astrophysics, Post  Bag 4, Ganeshkhind, Pune 411 007, India}
\author[0000-0001-7714-7076]{Anupreeta More}
\affiliation{Inter-University Centre for Astronomy and Astrophysics, Post  Bag 4, Ganeshkhind, Pune 411 007, India}
\affiliation{Kavli Institute for the Physics and Mathematics of the Universe (IPMU), 5-1-5 Kashiwanoha, Kashiwa-shi, Chiba 277-8583, Japan}
%%
%%%%%%%%%%%%%%%%%%%%%%%%%%%%%%%%%%%%%%%%%%%%%%%%%%%%%%%%%%%%%%%%%
%%%%%%%%%%%%%%%%%%%%%%%%%%%%%%%%%%%%%%%%%%%%%%%%%%%%%%%%%%%%%%%%%
\begin{abstract}
Identification and characterisation of Kilonovae (KNe) can be instrumental in improving our understanding of cosmology and astrophysics. However, their detection poses unique challenges due to rarity and faintness. Upcoming telescopes, with their deep imaging capabilities and wide field-of-views, will provide a unique opportunity to observe these rare and faint transients. The Rubin Legacy Survey of Space and Time (LSST) will generate a deluge of data, making it essential to deploy fast, efficient methods for identifying genuine KNe, especially when they are gravitationally lensed. To address this, we simulate realistic populations of both unlensed and lensed KNe in the six LSST bands. Comparing with the \typeonea\, Supernovae, we find that the KNe color evolution is more rapid and the two separate out when their colors are compared at two epochs. Since the mergers of compact binaries are probable progenitors of KNe, the KNe properties may be affected by the delay time distribution (DTD) of the mergers, which is dictated by the minimum delay time ($\delaytime$) and power-law slope. For longer $\delaytime$ and shallower slopes, we find an increased rate of detectable KNe in LSST. We generate the first statistically realistic lensed KNe population for different DTDs and find that the rate of detectable lensed KNe increases for DTDs with longer $\delaytime$ for a fixed slope. 
We further note that an \afterglow\,-like event at a redshift of 0.5~(1.0) needs magnification of at least 5~(44) to be detectable in LSST. 
\end{abstract}
%%
%%%%%%%%%%%%%%%%%%%%%%%%%%%%%%%%%%%%%%%%%%%%%%%%%%%%%%%%%%%%%%%%%
%%%%%%%%%%%%%%%%%%%%%%%%%%%%%%%%%%%%%%%%%%%%%%%%%%%%%%%%%%%%%%%%%
%%%%%%%%%%%%%%%%%%%%%%%%%%%%%%%%%%%%%%%%%%%%%%%%%%%%%%%%%%%%%%%%%%%%%%%%%%%%%%%%%%
%%%%%%%%%%%%%%%%%%%%%%%%%%%%%%%%%%%%%%%%%%%%%%%%%%%%%%%%%%%%%%%%%%%%%%%%%%%%%%%%%%
\section{Introduction} \label{S:intro}
With the detection of gravitational waves (GW) from the binary neutron star (BNS) event \knownbns\, and the optical counterpart \afterglow, the origins of Kilonovae (KNe) were firmly established \citep[e.g.,][]{coulter_2017,valenti2017,arcavi2017, Tanvir_2017,lipunov2017,soares_2017}. Numerous studies of KNe gave insights into the formation of heavy nuclei elements \citep[e.g.,][]{barnes_2013, kasen2013,tanaka_hotozaka_2013,tanaka_2018}, the physical structure and the mechanism that produced the electromagnetic emission at different wavelengths \citep[e.g.,][]{cowperthwaite_2017,siebert2017,nicholl_2017,villar_2017_dec}.
Since GW170817, there have been a few other GW events comprising neutron stars; however, none have had a confirmed detection of a KN counterpart. This indicates the challenges in conducting successful follow-up observations in the optical. The rarity and faintness of KNe make their discoveries extremely difficult. 

Occasionally, if some massive galaxy falls close to the line-of-sight between us and a source, the gravitational potential of the former can deflect and distort the emission from the source such that multiply lensed images of the source are formed. Additionally, the lensing magnification can cause distant transients such as Supernovae (SNe) and KNe to become detectable. Their lensing time delays ($\timedelay$) can offer a unique tool to constrain cosmological parameters such as the Hubble constant \citep[$\hnot$, e.g.,][]{kelly2015,rodney2021, Kelly2022, Chen2022, Frye2023, Pierel_2023_encore}. 
This is particularly important due to the growing Hubble tension arising due to a $5\sigma $ discrepancy between the early type  \citep[e.g.,][]{planck2014,planck2020,aiola_2020,bao2020,colas2020,dmico_2020,balkenhol2021,dutcher2021} and late type probes \citep[e.g.,][]{Sandage_2006_cepheids_sne, Freedman_2012_cepheid,riess_2016_sne_cepheids,riess_2018_cepheids,dhawan2018,riess_2019_cepheids, Pietrzynski2019, Riess_2021_cepheids}. The time-delay cosmography with lensed transients can play a crucial role as a one-step and independent probe of $\hnot$ \citep[e.g.,][]{treu_marshall_2016,oguri_2019_rev,liao_2022_rev}.
Lensed KNe systems have a similar advantage to lensed SNe, as both are transients. Moreover, the KNe will have GW counterparts, which can give additional constraints on the time delays and distances. Thus, providing improved constraints for time-delay cosmography.

Although the lensed KNe events are promising systems for $\hnot$ studies, it can be very challenging to detect them since lensing itself is rare, and coupled with the source being a faint transient. For example, the number of lensed SNe discovered to date is only a handful ( e.g.,` SN Refsdal': \citealt{kelly2015}; `SN Requiem': \citealt{rodney2021}; `AT2022riv': \citealt{Kelly2022}; `C22': \citealt{Chen2022}; `SN H0pe': \citealt{Frye2023}; `SN Encore': \citealt{Pierel_2023_encore}). Interestingly, if the lensed SN is indeed a \typeonea\, SN, then due to their standardizable candle nature, we can put better constraints on the lens models(`iPTF16geu': \citealt{more_iptf16geu_2017, goobar_2017}; `SN Zwicky': \citealt{goobar2023}). Even with the exciting discoveries of these lensed SNe, it is still hard to achieve the state-of-the-art precision level ($<$ 2 $\%$) of $\hnot$ estimates, due to insufficient data and very short lensing $\timedelay$s. However, new surveys such as the Rubin Legacy Survey of Space and Time (LSST)\citep{Ivezic_2019_lsst} and future missions such as Nancy Grace Roman Space Telescope \citep{Pierel_2021_roman} will increase the number of lensed SNe by an order of magnitude in the coming years \citep{oguri_2010, Goldstein_2017,wojtak_2019}. Furthermore, \citet{arendse2024} showed that, within 3~yr of observations of LSST, a `gold sample' of $\sim$ 10 lensed \typeonea\, SN (with $\timedelay>10$~days and SNR~$>$~5) per year can constrain $\hnot$ at a 1.5~$\%$ precision level (although also, see \citealt{schneider_2013}).

Even if there is no direct detection of lensed KNe to date, a few studies have predicted the detection of lensed GW in the upcoming fifth (O5) observing runs of the LIGO observatories and their optical counterparts in the wide and deep optical surveys such as the LSST. \citet{smith_2023} studied lensed KN detection focusing on the upcoming GW runs, and found that at least one lensed BNS merger per year will get detected in the O5 run. They also mentioned the lensed KNe will also be bright enough to get detected with LSST using deep Target of Opportunity (ToO) observations. However, the peak of the lensed KNe light curve will be at the fainter end of the LSST limit, depending on the lensing magnification. \citet{ma_lensed_kne_2023} showed that future GW detectors such as the Einstein telescope (Cosmic Explorer) can detect 5 (67) lensed BNS merger events per year. Although they predicted the detection of two electromagnetic counterparts per year with the JWST-like telescopes, the detection of these in Rubin LSST will be challenging. Here, we do not apply constraints from present or upcoming GW observatories a priori; we develop our methods focusing on the detectability of the optical counterparts in different LSST bands. For this reason, our lensed KNe sample is not limited by any redshift coming from GW detectors, which is crucial for lensing simulation as, in general, the lensed population peaks around higher source redshift ($\srcred$).    

For optimising KNe detections, \citet{Andreoni_2022} investigated follow-up strategies for ToO observations of GW candidates. In a different work, \citet{Andreoni_2021a} focused on the serendipitous searches of KNe events using a method developed for Zwicky Transient Facility \citep[ZTFReST:][]{Andreoni_2021b} based on time evolution of the light curves. In a recent study, using available LSST cadences, \citet{Andrade_2025} focus on finding the best combination of bands and frequency of visits of the same area in the sky to maximise the detection rate of KNe events. Here, we do not focus on the details of observing strategies to evaluate detectability. 
We refer to a KN as detectable if the peak of the light curve is brighter than the 30~sec single-exposure limit of the LSST Wide-Fast-Deep (WFD) survey. In some cases, we have also shown results for additional exposure times of 120~sec and 180~sec corresponding to possible deeper follow-ups with Rubin.  

In this work, we take into account different BNS merger rate distributions to generate a realistic KNe population for different binary stellar evolution parameters. When generating the KNe light curves, we consider a state-of-the-art three-component opacity model to include a more diverse light curve compared to \afterglow. We explain our KNe models in \autoref{S:kn_models}. We use real massive galaxies as foreground deflectors (or lenses) to generate the first statistically realistic lensed KNe population. We describe our methods to generate both unlensed and lensed populations in \autoref{S:methods}.  In \autoref{S:results}, we propose fast diagnostic metrics to quickly distinguish KNe from SNe. We also characterise the detectable KNe population properties with and without the lensing effects expected to be found in LSST and/or deeper Rubin data. Finally, we present a summary and our conclusions in \autoref{S:conclusion}. 
%%%%%%%%%%%%%%%%%%%%%%%%%%%%%%%%%%%%%%%%%%%%%%%%%%%%%%%%%%%%%%%%%%%%%%%%%%%%%%%%%%%
%%%%%%%%%%%%%%%%%%%%%%%%%%%%%%%%%%%%%%%%%%%%%%%%%%%%%%%%%%%%%%%%%%%%%%%%%%%%%%%%%%
\section{Kilonova models} \label{S:kn_models}
KNe light curve models and the BNS merger rate distribution with redshift are some of the main ingredients for generating the KNe population. Although there is a large uncertainty in some of these models, we use the most commonly and widely used models in this work and also consider the impact of variations in their parameters on the KNe population. Below, we describe the models for the merger rates and KNe light curves.
\subsection{BNS merger rates}\label{S:bns_merger}
For population studies of KNe, it is common to use a fixed BNS merger rate. For instance, while estimating the total number of KN events, \citet[][hereafter SC18]{scolnic_2018_kn_rate} used a fixed BNS merger rate of 1000 $\knrtgiga$. \citet{Andreoni_2019} also used a similar rate to study the effect of the LSST cadence on the detectability of the KNe population. However, in a more recent study, \citet{Andrade_2025} showed the importance of using a realistic BNS rate in determining the LSST cadence strategy on the detectability of the KNe population. They used an optimistic (pessimistic) rate of 295.7~(21.6)~$\knrtgiga$ with a median value at 105.5~$\knrtgiga$. 

In this study, we explore various BNS merger rate models, including a ``fixed" merger rate. We consider the formalism described in \citet[][hereafter \safKN]{saf_2019} 
to estimate the redshift-dependent merger rates for BNS. \safKN\ use the star formation rate (SFR) from \citet[hereafter MD14]{madau_dickinson_2014} and convolve it with a power-law type ($\propto\,(t - \delaytime)^{\alpha}$, where minimum delay time = $\delaytime$ and index of the power law = $\pwrlw$ ) delay-time distribution (DTD) to estimate the BNS merger rates. The two main parameters that control these distributions are $\delaytime$ and $\alpha$.
The DTD corresponds to the distribution of the evolution times between the epoch of formation of the neutron stars in a binary and the epoch of their final merger. 
For a small $\delaytime$ and steeper $\alpha$, we expect a quick merger of the BNSs, and the peak of the BNS merger distribution is likely to follow the peak of the SFR. 
For a large $\delaytime$ and shallower $\alpha$, we expect to have a delayed merger event, and hence, it will occur at low redshift (for more details, see Section 2 of \safKN). SAF19 also consider an efficiency factor ($\lambda$) of $10^{-5}~\rm{M}_{\odot}^{-1}$  which determines the fraction of stellar mass that will be converted to form BNS events.

%%%%%%%%%%%%%%%%%%%%%%%%%%%%%%%%%%%%%%%%%%%%%%%%%%%%%%%%%%%%%%%%%%%%%%%%%%%%%%%%%%
\subsection{KN light curves} \label{S:kn_light_curves}
Numerous studies on KN light curve models are found in the literature, along with publicly available packages to generate multi-band KN light curves. For instance, SC18 used a single spectral energy distribution (SED) model based on the \afterglow\, event to predict the detectable KNe events in future telescopes. In contrast, \citet{kasen_2017} and \citet{bulla_2019} used different parameters, such as the ejecta mass, the ejecta velocity and lanthanide fractions to create a large variety of KN light curves, which are not limited to the KN of GW170817. In another study, \citet{villar_2017_dec} produced KN light curves from a three-component model comprising 13 parameters, using publicly available \mosfit\ package \citep{metzger_2017, Villar_2017_nov}, and inferred the posterior distributions of their model parameters for the case of \afterglow. 

In this work, we use an open source \python-based software package \redback \footnote{\redbackdocs} for generating KNe light curves. Although \redback\, is mainly used for Bayesian inferences, we take advantage of its large and easily accessible library of different types of transients, such as KNe, SNe, and gamma-ray burst afterglows.  Among the many KN light curve models within \redback, we chose to use the \mosfit\ based three-component model. The list of the 13 free parameters and their functional form that we use to generate our KN light curve is given in \autoref{tab:all_kn_para_table}. In this model, each component has specific opacities by construction. For example, blue (red) ejecta has a typical opacity (\opacitysym\,) of $0.5~(10)$~\opacityunit\, which corresponds to the poor (rich) lanthanide fraction of that particular ejecta \citep{tanaka_2018}. 
Moreover, how fast the light curve will peak and evolve with time is also determined by ejecta properties present in a BNS merger event. For example, the slow evolution of light curves and the shift of the peak towards the redder wavelength are owing to ejecta with high opacities. 
%%%%%%%%%%%%%%%%%%%%%%%%%%%%%%%%%%%%%%%%%%%%%%%%%%%%%%%%%%%%%%%%%%%%%%%%%%%%%%
%\startlongtable
\tabcolsep7.0pt 
\begin{deluxetable}{lcc}
\tablecaption{List of model parameters to generate individual KN light curve. }
%\scriptsize
\tablehead{\colhead{Parameter Name}&\colhead{Function}&\colhead{Unit}
\label{tab:all_star_table}  
}
\startdata
$\mejone$, $\mejtwo$, $\mejthree$ & Uni (0.01, 0.03)&$\msun$\\ 
$\vejone$, $\vejtwo$ & Uni (0.1, 0.4)&c\\
$\vejthree$ & Uni (0.1, 0.7)&c\\
$\kappaone$ & $\mathcal{N}(\mu =0.5, \sigma=0.2,$ &$\mathrm{cm}^{2}/\mathrm{g}$\\
 & $,lo=0,hi=2)$ &\\
$\kappatwo$ & $\mathcal{N}(\mu =3, \sigma=1,$ &$\mathrm{cm}^{2}/\mathrm{g}$\\
 & $,lo=2,hi=5)$ &\\
 $\kappathree$ & $\mathcal{N}(\mu =10, \sigma=2,$ &$\mathrm{cm}^{2}/\mathrm{g}$\\
 & $,lo=5,hi=15)$ &\\
$T_{\mathrm{floor}1}$, $T_{\mathrm{floor}2}$ $T_{\mathrm{floor}3}$ & LogUni (100, 6000)&K\\
$\kappa_{\gamma}$ & Delta (peak=10)&$\mathrm{cm}^{2}/\mathrm{g}$\\
\enddata
\tablecomments{
\scriptsize The 13 parameters of the three-component model are used to generate the KN light curve.
Here, $M$, $v$, $\kappa$, and $T$ denote the masses, the velocities, the opacities, and the limiting temperature, respectively, for the three different types of ejecta. }
\label{tab:all_kn_para_table}
\end{deluxetable}
%%%%%%%%%%%%%%%%%%%%%%%%%%%%%%%%%%%%%%%%%%%%%%%%%%%%%%%%%%%%%%%%%%%%%%%%%%%%%%%%%%%%%%%
%%%
%%%%%%%%%%%%%%%%%%%%%%%%%%%%%%%%%%%%%%%%%%%%%%%%%%%%%%%%%%%%%%%%%%%%%%%%%%%%%%%%%%%%%%%
%%%%%%%%%%%%%%%%%%%%%%%%%%%%%%%%%%%%%%%%%%%%%%%%%%%%%%%%%%%%%%%%%%%%%%%%%%%%%%%%%%%%%%%
%%%%%%%%%%%%%%%%%%%%%%%%%%%%%%%%%%%%%%%%%%%%%%%%%%%%%%%%%%%%%%%%%%%%%%%%%%%%%%%%%%%%%%%
%%%%%%%%%%%%%%%%%%%%%%%%%%%%%%%%%%%%%%%%%%%%%%%%%%%%%%%%%%%%%%%%%%%%%%%%%%%%%%%%%%%%%%%
%%%%%%%%%%%%%%%%%%%%%%%%%%%%%%%%%%%%%%%%%%%%%%%%%%%%%%%%%%%%%%%%%%%%%%%%%%%%%%%%%%%%%%%
%%%%%%%%%%%%%%%%%%%%%%%%%%%%%%%%%%%%%%%%%%%%%%%%%%%%%%%%%%%%%%%%%%%%%%%%%%%%%%%%%%%%%%%
\section{Methodology}\label{S:methods}
In this section, we describe the underlying framework for generating the unlensed and lensed KN populations. For the generations of the KNe population, we use several different values of $\delaytime$ and $\pwrlw$ and incorporate the $\srcred$ dependency in the BNS merger rate. Then, we attach a light curve to each of these KNe events using the \mosfit\, three-component model. Finally, we use these KNe as sources and massive galaxies as lenses to create a realistic sample of KNe with lensed images. Below, we describe each of these processes in detail.
\subsection{Generation of the unlensed KN population}\label{S:unlensed_kn_pop}
As described in \autoref{S:bns_merger}, we start with the SAF19 model for the BNS merger rate. Accurate values of the $\delaytime$ and $\alpha$ of the delay time distribution are not known from the current gravitational wave detectors, unfortunately. Hence, following the SAF19 formalism, we take combinations of different $\delaytime=$~[10~Myr, 100~Myr, 1~Gyr] and $\alpha=$~[-1.5, -1.0, -0.5] for this study.
%%%%%%%%%%%%%%%%%%%%%%%%%%%%%%%%%%%%%%%%%%%%%%%%%%%%%%%%%%%%%%%%%%%%%%%%%%%%%%%%%%%%
%%%%%%%%%%%%%%%%%%%%%%%%%%%%%%%%%%%%%%%%%%%%%%%%%%%%%%%%%%%%%%%%%%%%%%%%%%%%%%%%%%%%
\begin{figure}
\gridline{\fig{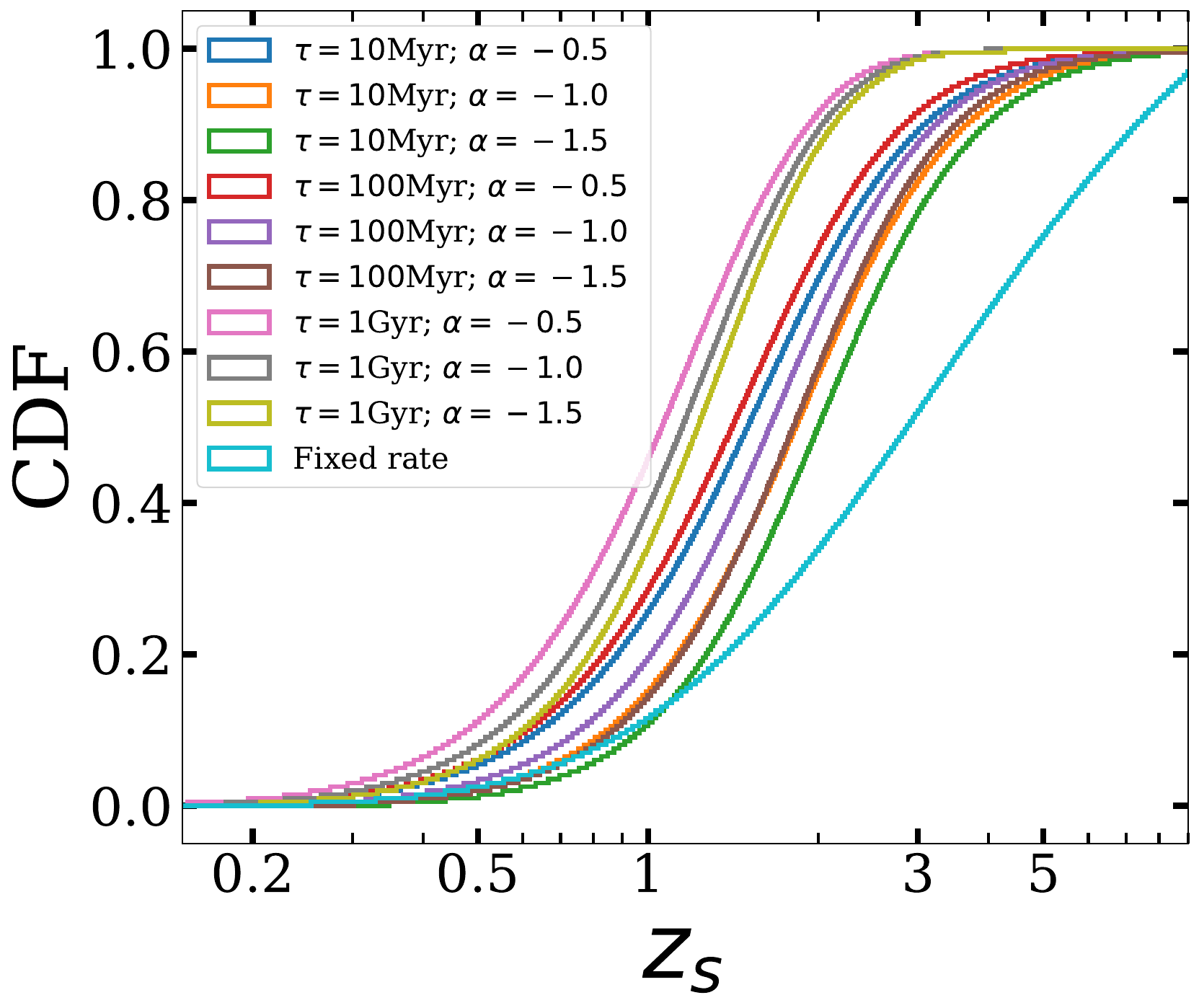}{0.45\textwidth}{}
          }
\caption{Cumulative density function (CDF) of BNS merger events with KNe redshifts ($\srcred$). The colors correspond to different sets of delay time ($\delaytime$) and power-law index ($\pwrlw$) from \safKN\, and a fixed value of 1000~$\knrtgiga$. 
The distribution with the longest $\delaytime$ peaks at low redshifts. 
For a fixed $\delaytime$, the distributions with shallower (steeper) slope peak at higher (lower) redshifts. The ``fixed rate" distribution peaks at much higher redshift and has a longer tail compared to the other distributions.}
\label{fig:bns_merger_rate_saf19}
\end{figure}
%%%%%%%%%%%%%%%%%%%%%%%%%%%%%%%%%%%%%%%%%%%%%%%%%%%%%%%%%%%%%%%%%%%%%%%%%%%%%%%%%%%%
%%%%%%%%%%%%%%%%%%%%%%%%%%%%%%%%%%%%%%%%%%%%%%%%%%%%%%%%%%%%%%%%%%%%%%%%%%%%%%%%%%%%
In \autoref{fig:bns_merger_rate_saf19}, we show how the cumulative merger rate distributions of KNe change with $\srcred$, for varying ($\delaytime$, $\alpha$). 
We notice the cumulative density functions (CDFs) corresponding to $\delaytime=1~\rm{Gyr}$ for all of the $\pwrlw$ values show a peak at lower $\srcred$ compared to the other distributions. 
For a given $\delaytime$, distributions with low $\pwrlw$ (i.e. steep negative slope) show a peak towards the higher $\srcred$. This is due to the fact that high negative $\pwrlw$ values create BNS distributions which merge quickly after the formation of their progenitors. In the low $\srcred$ regime, the $\delaytime$ of 1~Gyr mostly dominates the population of the BNS mergers. However, in the higher $\srcred$ it is harder to distinguish which combination of $\delaytime$ and $\pwrlw$ is dominating due to their degeneracies. Nevertheless, the fixed rate distribution peaks at a much higher $\srcred$ compared to all other distributions, and this also shows a longer tail in the $\srcred$ distributions.
As the peak and the profile of the BNS merger rate distribution depend on the $\delaytime$ and $\pwrlw$, we consider a range of values that envelops the highest and lowest BNS merger rates. 
We refer to each set of ($\delaytime$,~$\pwrlw$) as a scenario and repeat the entire analysis for each of these cases. We randomly draw KN sources from the merger rate distributions mentioned above and use them as sources in our (un)lensing simulation.

\begin{figure}
\gridline{\fig{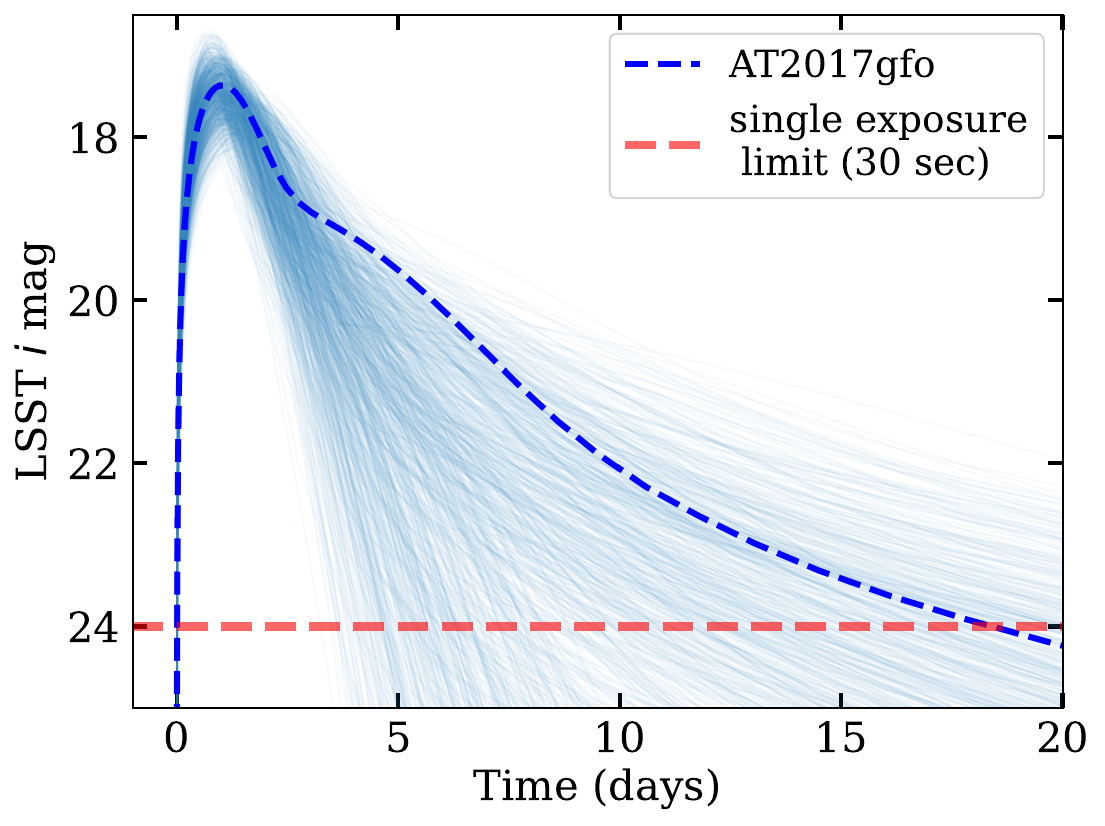}{0.47\textwidth}{}}
\caption{Light curves for a population of 1000 KNe in the LSST $i$ band (blue thin curves), where the physical parameters are drawn from the distributions described in \autoref{tab:all_kn_para_table}. The dashed blue curve shows the AT2017gfo model. The horizontal red dashed line shows the single-exposure limit (30~sec) for the LSST $i$ band.} 
\label{fig:mosfit_kn_modles}
\end{figure}

After fixing the $\srcred$ of a particular KN, the next step is to attach a light curve to that KN. For this, we use the \mosfit\ package described in \autoref{S:kn_light_curves}.
We randomly draw the masses of three types of ejecta from a uniform distribution with lower (upper) bounds at 0.01~(0.03)~$\msun$. The velocity of the blue and purple ejecta are drawn from a uniform distribution with lower (upper) bounds at 0.1~$c$, whereas the red ejecta velocity values are from a uniform distribution with lower (upper) bounds at 0.1~(0.7)~$c$. The three $\kappa$ values are taken from three truncated normal distributions with mean values at 0.5, 3, and 10 with sigma values of 0.2, 1, and 2, respectively. The lower (upper) bounds of the three normal distributions are 0~(2), 2~(5), and 5~(15) for blue, purple, and red ejecta, respectively. We mention all these parameters and their distributions in \autoref{tab:all_kn_para_table}.
In \citet{villar_2017_dec}, there is a strong correlation of ejecta masses ($\mejone$, $\mejtwo$, $\mejthree$) with the velocities ($\vejone$, $\vejtwo$, $\vejthree$) for all three types of ejecta in case of \afterglow. This correlation can increase the peak brightness of the light curve as both high ejecta mass and velocity lead to brighter light curves. However, this correlation may not hold for the entire KNe population; we do not apply any mass-velocity correlation in this study. 
In \autoref{fig:mosfit_kn_modles}, the thin blue curves are generated using parameter values drawn from their distributions given in \autoref{tab:all_kn_para_table}, which is $\srcred$ = $0.0098$. We use AT2017gfo redshift to put all these light curves in the same $\srcred$ for comparison. For reference, we also show the light curve from \citet{villar_2017_dec} for AT2017gfo with a thick blue dashed curve.
%%%%%%%%%%%%%%%%%%%%%%%%%%%%%%%%%%%%%%%%%%%%%%%%%%%%%%%%%%%%%%%%%%%%%%%%%%%%%%%%%%%%%
%%%%%%%%%%%%%%%%%%%%%%%%%%%%%%%%%%%%%%%%%%%%%%%%%%%%%%%%%%%%%%%%%%%%%%%%%%%%%%%%%%%%%
Next, we use the KNe redshifts to determine their apparent magnitudes in each of the LSST bands. Our methodology ensures the generation of a more diverse sample of the KNe population rather than being limited only by the AT2017gfo light curve models. In our simulations, if the peak magnitude of a transient is brighter than the LSST $r$ and $i$ band 30~sec single-exposure limit, we call that a bright transient detectable in the LSST.
\subsection{Generation of the lensed KN population}\label{S:lens_pop}
To generate the lensed KN population, we use the \texttt{simct} \citep{more_simct2016} python package, which is designed for generating statistically realistic hybrid mock lens population. Since \texttt{simct} includes lensed sources such as galaxies and quasars, we adapt the code to generate the lensed transient population, such as KNe (also, see \cite{prajakta2025}). It makes use of \texttt{glafic} \citep{oguri_2010} to solve lensing equations.
 
For the foreground lens population, we use galaxies from the latest third data release of Hyper Suprime-Cam (HSC) galaxy survey data \citep{hsc_pdr3}. 
We consider galaxies with stellar mass $>5\times10^{10}\msun$ and the ratio of the SFR-to-stellar mass (in $\msun$) to be less than $1\times10^{-10}$ per year. Using these selection criteria, we construct the sample of massive early-type galaxies from the HSC data. We assume the Singular Isothermal Ellipsoid (SIE) model \citep{kormann1994, Koopmans_2009} for the mass distribution of the lens galaxies. The SIE model parameters include the velocity dispersion ($\sigmavd$), ellipticity ($e$), position angle (PA) and position ($x$,$y$) of the lens mass distribution. We determine the SIE model mass parameters assuming mass follows light.
As a result, we begin by obtaining the light properties of the (lens) galaxies such as the redshift ($\lensred$), magnitudes in $g$, $r$ and $i$ bands, shape parameters ($e$, PA) from the HSC galaxy catalogue. Then we use the luminosity (L) to estimate lens $\sigmavd$ by following the $L-\sigma$ relation of \citet{parker_2007}. We draw shear randomly from a uniform distribution (0.0,0.2), respectively, consistent with studies in the literature \citep[e.g.,][]{Wong_2011}. 

We obtain the redshift distribution of the background KNe for three choices of $\delaytime$ and $\pwrlw$ of the DTDs.
For every potential lens galaxy from the HSC catalogue, we draw a KN from the given redshift distribution and determine the lensing probability for this system to be a lens based on the lensing cross-section.
%%%%%%%%%%%%%%%%%%%%%%%%%%%%%%%%%%%%%%%%%%%%%%%%%%%%%%%%%%%%%%%%%%%%%%%%%%%%%%%%
%%%%%%%%%%%%%%%%%%%%%%%%%%%%%%%%%%%%%%%%%%%%%%%%%%%%%%%%%%%%%%%%%%%%%%%%%%%%%%%%%
\begin{figure*}
\gridline{\fig{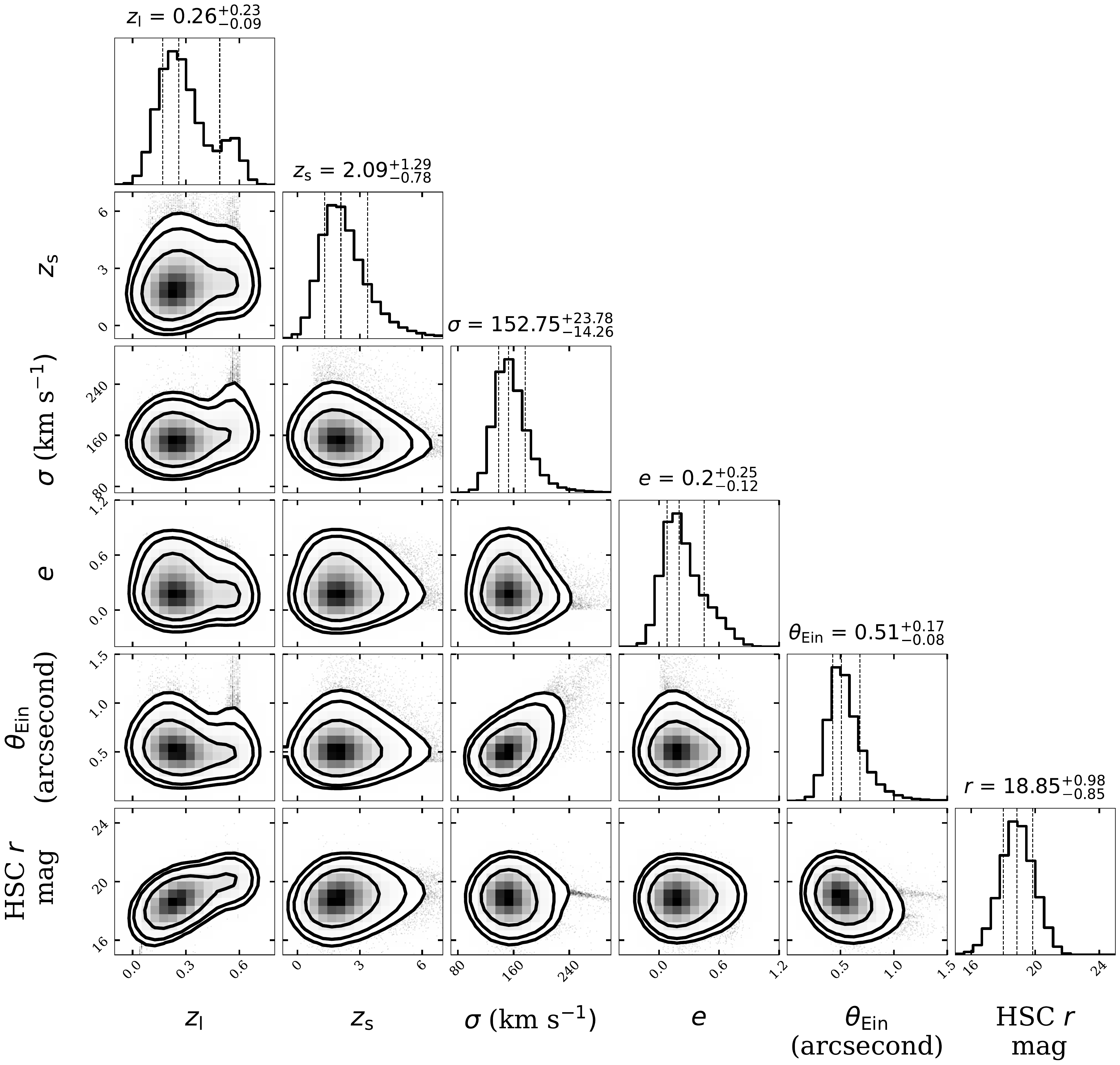}{0.90\textwidth}{}}
\caption{Distributions of properties of the simulated lens population generated using real HSC galaxies. From left, we show the lens redshift ($\lensred$), source redshift ($\srcred$), lens velocity dispersion ($\sigmavd$), ellipticity ($e$), Einstein radius ($\einrad$), and magnitude (HSC $r$ band) of the lens galaxy. In the two dimensional joint distributions, different contour levels show $68 \%$, $90 \%$, and $95 \%$ enclosed probability region. Vertical lines in one dimensional distributions show the median, $16 \%$, and $84 \%$ confidence intervals. These distributions are representative of typical galaxy-scale lens samples. } 
\label{fig:velocity_dis_lens}
\end{figure*}
%%%%%%%%%%%%%%%%%%%%%%%%%%%%%%%%%%%%%%%%%%%%%%%%%%%%%%%%%%%%%%%%%%%%%%%%%%%%%%%%%
%%%%%%%%%%%%%%%%%%%%%%%%%%%%%%%%%%%%%%%%%%%%%%%%%%%%%%%%%%%%%%%%%%%%%%%%%%%%%%%%%
In \autoref{fig:velocity_dis_lens}, we show the redshift ($\srcred$, $\lensred$) distributions of both lens and sources, $\sigmavd$, $e$ of the lens,  the $\einrad$ and the $r$ magnitude of the lens. The median values of $\lensred$, $\srcred$, $\sigmavd$ , $e$, $\einrad$, and $r$ mag are 0.26, 2.09, 0.2, 152~$\velunit$, 0.51~arcsec, and 18.9~mag, respectively. The statistical properties are consistent with expected properties of galaxy-scale lens population. The KNe population used here corresponds to $\delaytime=100~\rm{Myr}$ and $\alpha=-1.5$ of the DTD.

\begin{equation}
\tau = \frac{\sigmalens}{4\pi}
\label{eq:optical_depth}
\end{equation}
where the $\sigmalens$ is the solid angle formed by the lens deflector. The lens galaxy need to fall within this $\sigmalens$ in order to create multiple images of the source. So, after all these steps, we can create a population of lensed KNe with multiple images. To create a sufficiently large enough sample of lensed KNe sample, we use a boost factor of $10^{20}$.  

\section{Results}\label{S:results}
In this section, we describe the results from the analyses of both the unlensed and the lensed KNe populations. Although the focus of this study is the discovery of lensed KNe in the LSST data, we start our analysis with the identification of the unlensed KNe population for a few reasons. As the lensed KNe are much rarer, it is useful to develop the tools to find the more common unlensed KNe, which may then also apply to the lensed KNe. Hence, we analyse the population of unlensed KNe for varied BNS merger rate distributions and then test their detectability in the LSST. We perform similar analyses on the lensed KNe.

\subsection{Unlensed KNe: Properties and Detectability}\label{S:bright_pop_detect}
For a set of different $\delaytime$ and $\alpha$ (see \autoref{tab:kn_rate_table}), we simulate the KNe population following several BNS merger rate distributions (see \autoref{fig:bns_merger_rate_saf19}). We determine the fraction of KNe that will be detectable in the $i$ and $r$ bands of LSST. In \autoref{tab:kn_rate_table}, we mention the rate of detectable KNe for different $\delaytime$, $\alpha$, and single-exposure time-limits of LSST. For a fixed $\delaytime$, we observe the rate of KNe increases with steeper $\alpha$. The reason for this is that $\alpha$ controls the slope of the BNS merger rate distributions, and a steeper negative slope allows the binary neutron stars to merge quickly and follow the MD14 SFR distributions closely. As a result, we get a larger rate of KNe. However, for a steeper slope, the KNe will also be at higher redshifts compared to a shallower slope, as the former population will merge quickly after the star formation. This directly affects the rate of detectable KNe in the LSST. As the slope becomes steeper, the percentage of detectable KNe decreases for all exposure limits. Now, for a fixed $\alpha$, the $\delaytime$ also affects the detectable KNe population. If the $\delaytime$ is long, for example, 1~Gyr, we will have fewer total KNe compared to a $\delaytime = 10~\rm{Myr}$, as the long $\delaytime$ will increase the time gap between the merger of BNS events from the formation of the progenitors. However, for a long $\delaytime$, the KNe population will peak at a lower $\srcred$ compared to that for a short $\delaytime$. This also reflects in the detectability rate of the KNe in the LSST. 

\begin{figure*}
\gridline{\fig{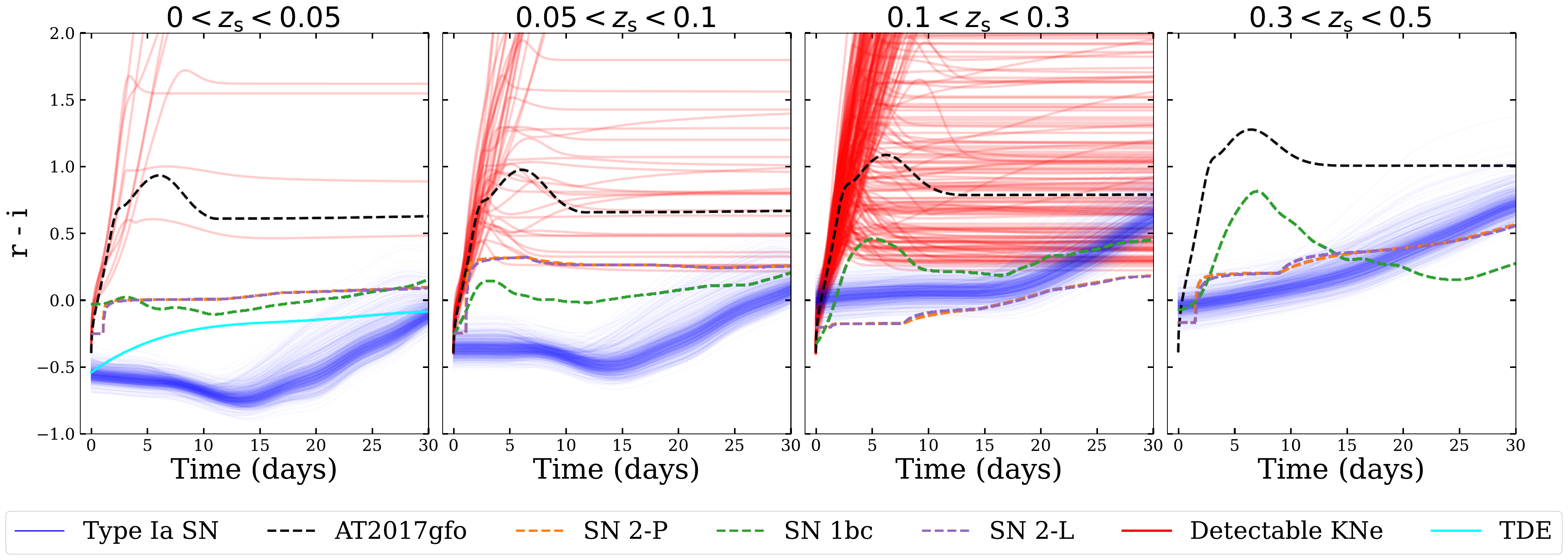}{0.94\textwidth}{}}
\caption{Color ($r-i$) evolution as a function of time for various types of transients.
The red and blue curves correspond to the detectable KNe and \typeonea\, SNe, respectively. 
Most of the detectable KNe peaks around $0.1 <\srcred < 0.3$ in the LSST.
Dashed curves in black, orange, green, and violet show the evolution of AT2017gfo, Type SN 2-P, Type SN 1bc, AND Type SN 2-L, respectively.} 
\label{fig:col_mag_evolution_plot}
\end{figure*}

In \autoref{tab:kn_rate_table}, for a fixed $\pwrlw$, we note that the detectable KNe fraction increases by more than a factor of 2 for two different $\delaytime=$ 10~Myr and 1~Gyr, for all three exposure limits. Finally, we estimate the rate of detectable KNe for a ``fixed merger" rate distribution of 1000 $\knrt$. While the absolute rate of detectable KNe is higher in the ``fixed rate" distribution compared to others, the fraction of detectable KNe is comparable.
%%%%%%%%%%%%%%%%%%%%%%%%%%%%%%%%%%%%%%%%%%%%%%%%%%%%%%%%%%%%%%%%%%%%%%%%%%%%%%
%%%%%%%%%%%%%%%%%%%%%%%%%%%%%%%%%%%%%%%%%%%%%%%%%%%%%%%%%%%%%%%%%%%%%%%%%%%%%%
\startlongtable
\tabcolsep9.0pt 
\begin{deluxetable*}{cccccc}
\tablecaption{KNe population properties for varied BNS merger rate distributions }
\scriptsize
\tablehead{\colhead{Minimum delay time }&\colhead{Power-law Slope}&\colhead{Total population }
&\multicolumn{3}{c}{Detectable Population ($\rm yr^{-1}$)(\% of Total population)}\\
\colhead{$\delaytime$}&\colhead{$\pwrlw$}&\colhead{within $\srcred=10.0(0.5)$}
&\multicolumn{3}{c}{Exposure time }\\
\colhead{}&\colhead{}&\colhead{ $\times 10^{5}$ $\rm yr^{-1}$}
&\colhead{30~sec}
&\colhead{120~sec}&\colhead{180~sec}
}
\startdata
10~Myr& $-0.5$&2.22(0.12)&816(0.37)&1935(0.87)&2372(1.07)\\
10~Myr& $-1.0$&3.71(0.09)&504(0.14)&1231(0.33)&1548(0.42)\\
10~Myr& $-1.5$&4.70(0.07)&315(0.07)&817(0.17)&1023(0.22)\\
\hline
100~Myr& $-0.5$&2.01(0.13)&840(0.42)&2027(1.01)&2484(1.24)\\
100~Myr& $-1.0$&3.05(0.11)&574(0.19)&1432(0.47)&1798(0.59)\\
100~Myr& $-1.5$&3.95(0.09)&404(0.10)&1048(0.27)&1317(0.33)\\
\hline
1~Gyr& $-0.5$&1.26(0.14)&1008(0.80)&2349(1.86)&2851(2.26)\\
1~Gyr& $-1.0$&1.63(0.14)&833(0.51)&2012(1.23)&2505(1.54)\\
1~Gyr& $-1.5$&2.00(0.13)&689(0.34)&1682(0.84)&2087(1.04)\\ 
\hline
Fixed Rate&Fixed Rate&7.43(0.20)&1485(0.20)&3418(0.46)&4212(0.57)\\
\enddata
\tablecomments{
\scriptsize Rate of unlensed and detectable KNe candidates for different combinations of $\delaytime$ and $\pwrlw$ of the DTD resulting in BNS merger rate distributions. Here we use three different single-exposure limits of 30~sec, 120~sec, and 180~sec for LSST bands.}
\label{tab:kn_rate_table}
\end{deluxetable*}
%%%%%%%%%%%%%%%%%%%%%%%%%%%%%%%%%%%%%%%%%%%%%%%%%%%%%%%%%%%%%%%%%%%%%%%%%%%%%%%%%%%
%
Now, it is important to be able to distinguish the KNe from other transients found in LSST for which we use multi-band and multi-epoch LSST data. In \autoref{fig:col_mag_evolution_plot}, we compare the color evolution of different transients such as Type I(a, bc) SNe, Type II-(P, L) SNe and tidal disruption event (TDE) with KNe population. Dashed black curve shows the time evolution of the light curve of \afterglow\ using parameter values from \citet{villar_2017_dec}, in LSST bands, where the $\srcred$ is the mean value of the two $\srcred$ end of that particular bin. In the first panel of \autoref{fig:col_mag_evolution_plot}, we add the color evolution of a very nearby (at 66 Mpc) tidal disruption event (TDE) AT2019qiz \citep{nicholl_tde_2020}. To estimate the color evolution of the TDE, we interpolated the ZTF $i$ and $r$ band magnitude for this event, as the ZTF $i$ and $r$ band central wavelengths are very similar to those of LSST. For the generation of the light curve of \typeonea\, SNe population, we closely follow the formalism of \citet{mane_2024} (see Section 2.3 in \citet{mane_2024}).
%%%%%%%%%%%%%%%%%%%%%%%%%%%%%%%%%%%%%%%%%%%%%%%%%%%%%%%%%%%%%%%%%%%%%%%%%%%%%%%%%%%
%%%%%%%%%%%%%%%%%%%%%%%%%%%%%%%%%%%%%%%%%%%%%%%%%%%%%%%%%%%%%%%%%%%%%%%%%%%%%%%%%%%
%%%%%%%%%%%%%%%%%%%%%%%%%%%%%%%%%%%%%%%%%%%%%%%%%%%%%%%%%%%%%%%%%%%%%%%%%%%%%%%%%%%
%%%%%%%%%%%%%%%%%%%%%%%%%%%%%%%%%%%%%%%%%%%%%%%%%%%%%%%%%%%%%%%%%%%%%%%%%%%%%%%%%%%
In \autoref{fig:col_mag_evolution_plot}, we show the color evolution of the detectable KNe (\typeonea\, SNe) sample using red (blue) colored thin curves in LSST $r$ and $i$ bands. From the figure, it is evident that the color evolution of detectable KNe is much faster than the other transients, especially compared to the \typeonea\, SNe. Different panels of \autoref{fig:col_mag_evolution_plot}, show different $\srcred$ bins. The rate of detectable KNe increases with increasing $\srcred$, due to the fact of covering a larger volume. However, in the $0.3 < \srcred < 0.5$, we do not find any KNe as their brightness falls below the detectable limit of LSST $r$ and $i$ bands. From the different $\srcred$ bins and the evolution time scale, it is quite evident that we can classify these two populations using colors around the peak magnitude in the low $\srcred$ regime.

\begin{figure*}
\gridline{\fig{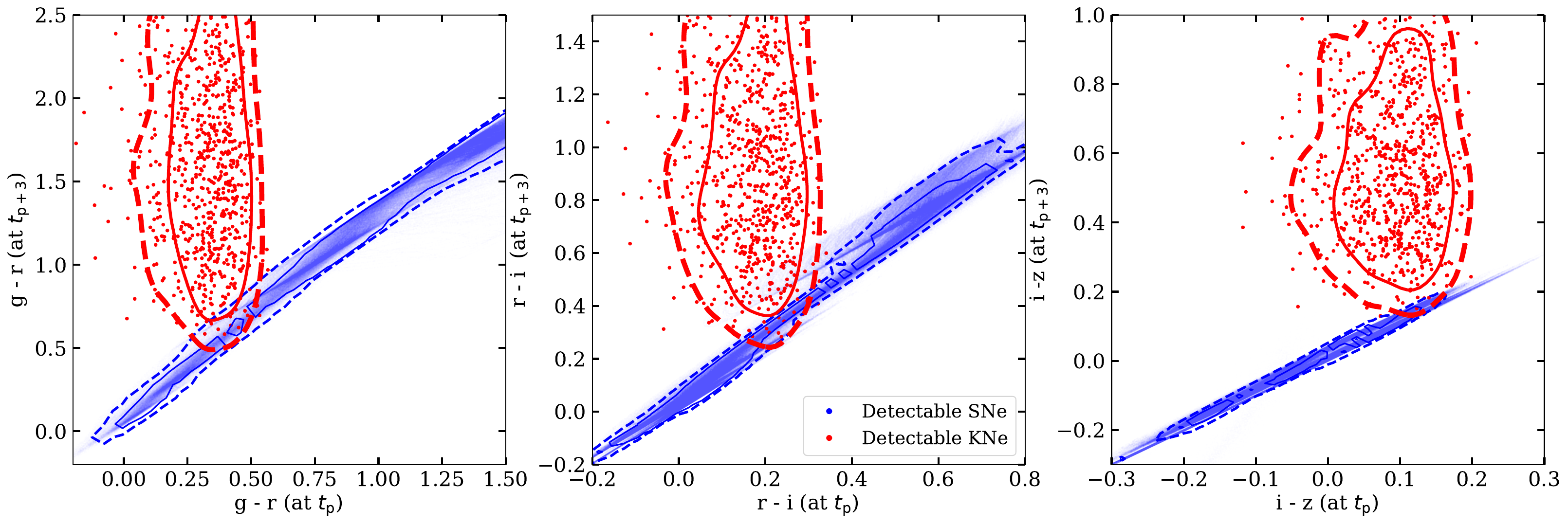}{0.90\textwidth}{}}
\caption{Comparison of colors at $\tpeak$ with the color at $\tpeakthree$ for the population of detectable KNe (red) and Type Ia SNe (blue) in LSST. Different panels show different color combination of the LSST bands. Solid (dashed) contours show the 68 (95) $\%$ confidence levels.}
\label{fig:int_sn_kn_color_color}
\end{figure*}

\begin{figure}
\gridline{\fig{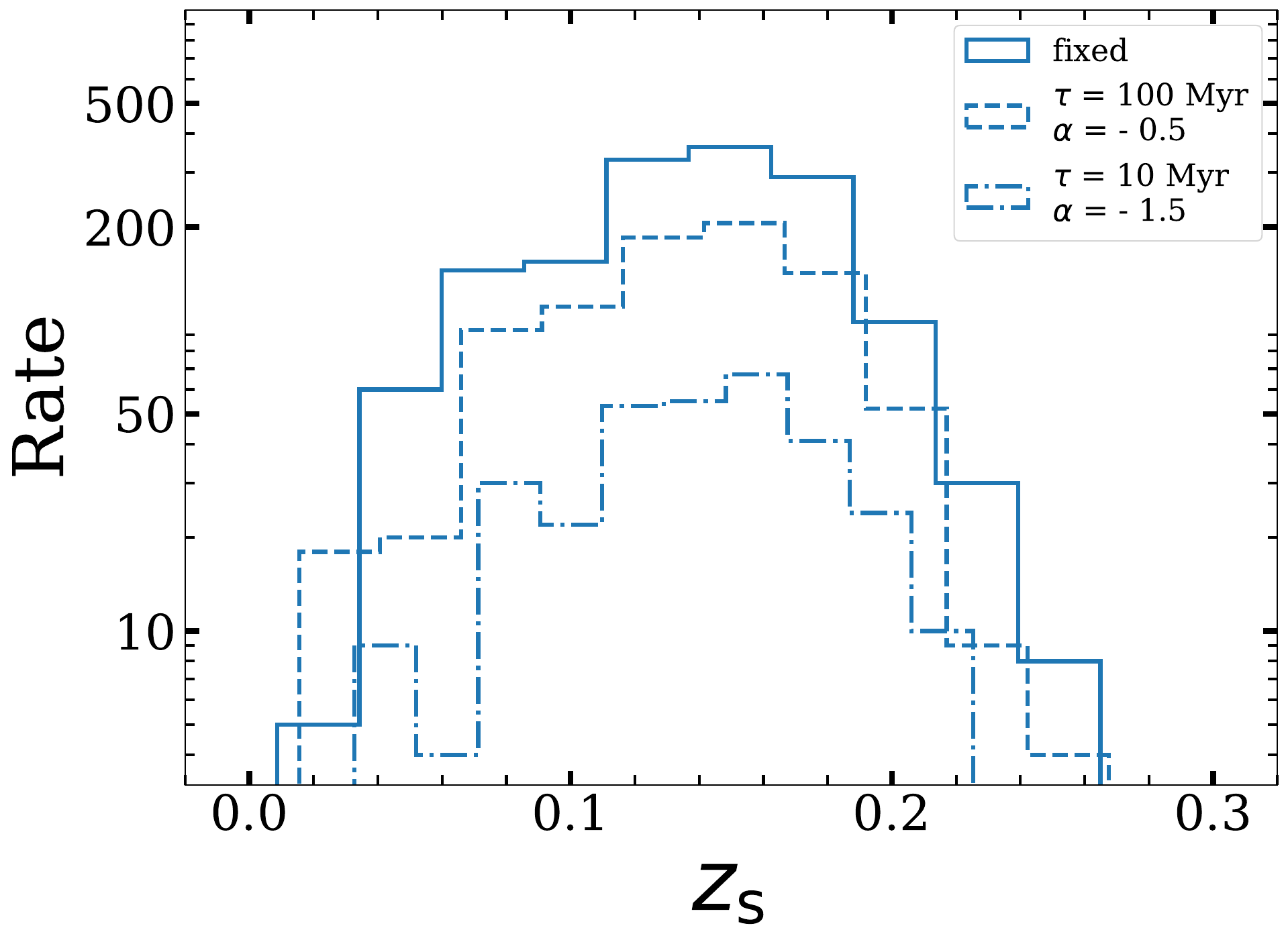}{0.45\textwidth}{}}
\caption{Redshift ($\srcred$) distribution of detectable KNe rate for three different types of scenarios. The $\delaytime = 10~\rm{Myr}$ and $\pwrlw = -1.5$ show the minimum rate of detectable KNe population. We show the ``fixed rate" population for reference, which has the maximum rate of detectable KNe. We find that all of the unlensed detectable KNe are likely to arise from $\srcred \lesssim 0.3$ in the LSST.}
\label{fig:redshift_dist_bright_kne}
\end{figure}

\begin{figure}
\gridline{\fig{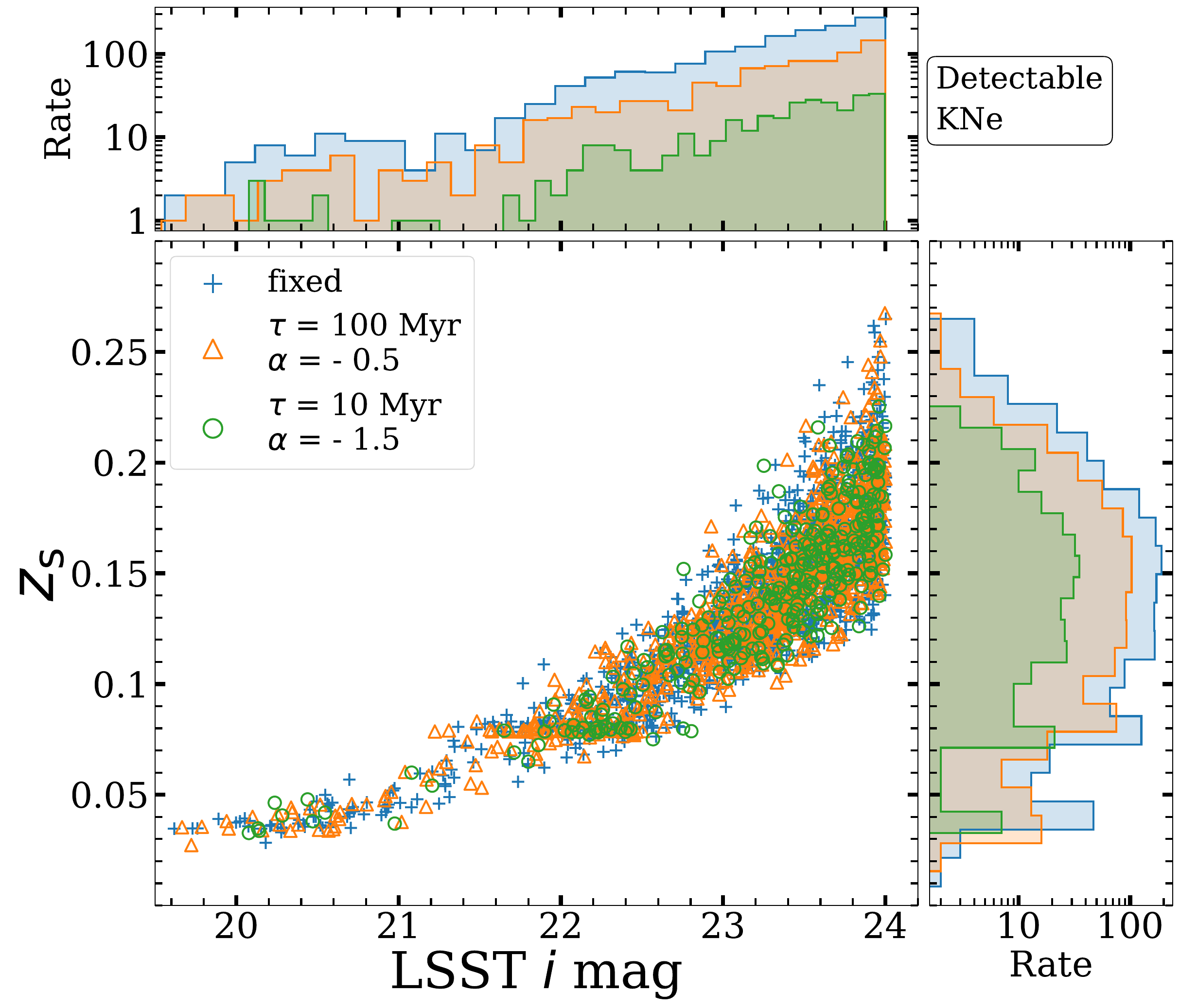}{0.45\textwidth}{}}
\caption{Redshift and LSST $i$ band distribution of detectable KNe for three different scenarios. We depict these three scenarios with different colored markers. Although most of the KNe come from the faint end, there are some bright nearby KNe ($i\sim20$ and $\srcred \sim 0.05$).} 
\label{fig:redshift_mag_dist}
\end{figure}

Now, to create uniformity among the time evolution of different transients, we prefer to choose the time ($\tpeak$) at the peak brightness and the time ($\tpeakthree$) after three days of evolution, for both KNe and \typeonea\, SNe, in different LSST bands. In \autoref{fig:int_sn_kn_color_color}, we show the transient color at $\tpeak$ of the light curve with the color at $\tpeakthree$ in different LSST bands for both detectable \typeonea\, SNe and KNe. We depict the detectable SNe population using blue dots, and all red dots denote the detectable KNe population. Solid (dashed) contours show the $68~(95)\%$ boundary of the populations. From this illustration, we notice the detectable KNe and SNe constitute two different regions, specifically in the $g$, $r$, $i$, and $z$ combinations of LSST bands. Hence, we find the combination of $i$, $z$ and $g$, $r$ can create a good set of bands to easily distinguish KNe from the SNe. Nevertheless, just being brighter than the 30~sec or 180~sec single-exposure limit of LSST bands will not ensure detection of KN. For this, we need to deploy several LSST observing strategies which are beyond the scope of this work.
%%%%%%%%%%%%%%%%%%%%%%%%%%%%%%%%%%%%%%%%%%%%%%%%%%%%%%%%%%%%%%%%%%%%%%%%%%
%%%%%%%%%%%%%%%%%%%%%%%%%%%%%%%%%%%%%%%%%%%%%%%%%%%%%%%%%%%%%%%%%%%%%%%%%%
%%%%%%%%%%%%%%%%%%%%%%%%%%%%%%%%%%%%%%%%%%%%%%%%%%%%%%%%%%%%%%%%%%%%%%%%%%
%%%%%%%%%%%%%%%%%%%%%%%%%%%%%%%%%%%%%%%%%%%%%%%%%%%%%%%%%%%%%%%%%%%%%%%%%%
%%%%%%%%%%%%%%%%%%%%%%%%%%%%%%%%%%%%%%%%%%%%%%%%%%%%%%%%%%%%%%%%%%%%%%%%%%
Next, we study the $\srcred$ distribution of the detectable KNe population. Here we choose three different combinations of $\delaytime$ and $\pwrlw$ from DTDs to represent the high, medium, and low rates of detectable KNe population (see \autoref{fig:redshift_dist_bright_kne}). The highest rate comes from the ``fixed" rate distribution, and that decreases with $\delaytime = 100~(10)~\rm{Myr}$ and $\pwrlw = -0.5~(-1.5)$. For all three populations, the peaks of the distributions are at $\srcred\sim0.15$. LSST can observe these detectable KNe populations up to a maximum $\srcred$ of 0.25. We investigate how the LSST $i$ band peak magnitude changes with $\srcred$ for these three scenarios. In \autoref{fig:redshift_mag_dist}, all the colored circles are the detectable KNe from different $\delaytime$ and $\pwrlw$, and again the ``fixed" population have the highest rate compared to the other two populations. Although the majority of these KNe are coming from the fainter end, there are a handful of KNe that have a peak magnitude as bright as 20 in the $i$ band.
%%%%%%%%%%%%%%%%%%%%%%%%%%%%%%%%%%%%%%%%%%%%%%%%%%%%%%%%%%%%%%%%%%%%%%%%%%%%%%%
\subsection{Lensed KNe: Properties and Detectability}
\label{S:lens_img_prop}
Here, we focus on the lensed KNe properties. As the lensing calculations are computationally time-intensive, we limit our lensed samples to three main scenarios. For these three scenarios, we fix the $\pwrlw$ to $-1.5$ and consider three choices for the $\delaytime$, namely, 10~Myr, 100~Myr, and 1~Gyr. Although these choices produce a small number of unlensed KNe at low redshifts, they generate a larger number of KNe at relatively high redshifts. Hence, the three chosen scenarios are optimal for lensing. For all of the lensing simulations, we select massive early-type galaxies from the HSC galaxy catalogue as the lensing deflector populations (described in \autoref{S:lens_pop}). We follow the \citet{a_more2022} formalism closely for generating lensed population properties. We only retain those doubly or quadruply imaged lenses for which the $\einrad$ is larger than 0.4~arcsec.
%%%%%%%%%%%%%%%%%%%%%%%%%%%%%%%%%%%%%%%%%%%%%%%%%%%%%%%%%%%%%%%%%%%%%%%%%%%%%%%%%
%%%%%%%%%%%%%%%%%%%%%%%%%%%%%%%%%%%%%%%%%%%%%%%%%%%%%%%%%%%%%%%%%%%%%%%%%%%%%%%%%
%%%%%%%%%%%%%%%%%%%%%%%%%%%%%%%%%%%%%%%%%%%%%%%%%%%%%%%%%%%%%%%%%%%%%%%%%%%%%%%%%
This constraint is based on the best possible seeing of the LSST and is chosen conservatively to imply that we are unlikely to resolve multiple images if they are within 0.4~arcsec of each other. 
In \autoref{tab:lensed_kn_rate_table}, we list the fraction of lensed KNe for all three different scenarios.
%%%%%%%%%%%%%%%%%%%%%%%%%%%%%%%%%%%%%%%%%%%%%%%%%%%%%%%%%%%
%%%%%%%%%%%%%%%%%%%%%%%%%%%%%%%%%%%%%%%%%%%%%%%%%%%%%%%%%%%
% %%%%%%%%%%%%%%%%%%%%%%%%%%%%%%%%%%%%%%%%%%%%%%%%%%%%%%%%%%%
\begin{figure*}
\gridline{\fig{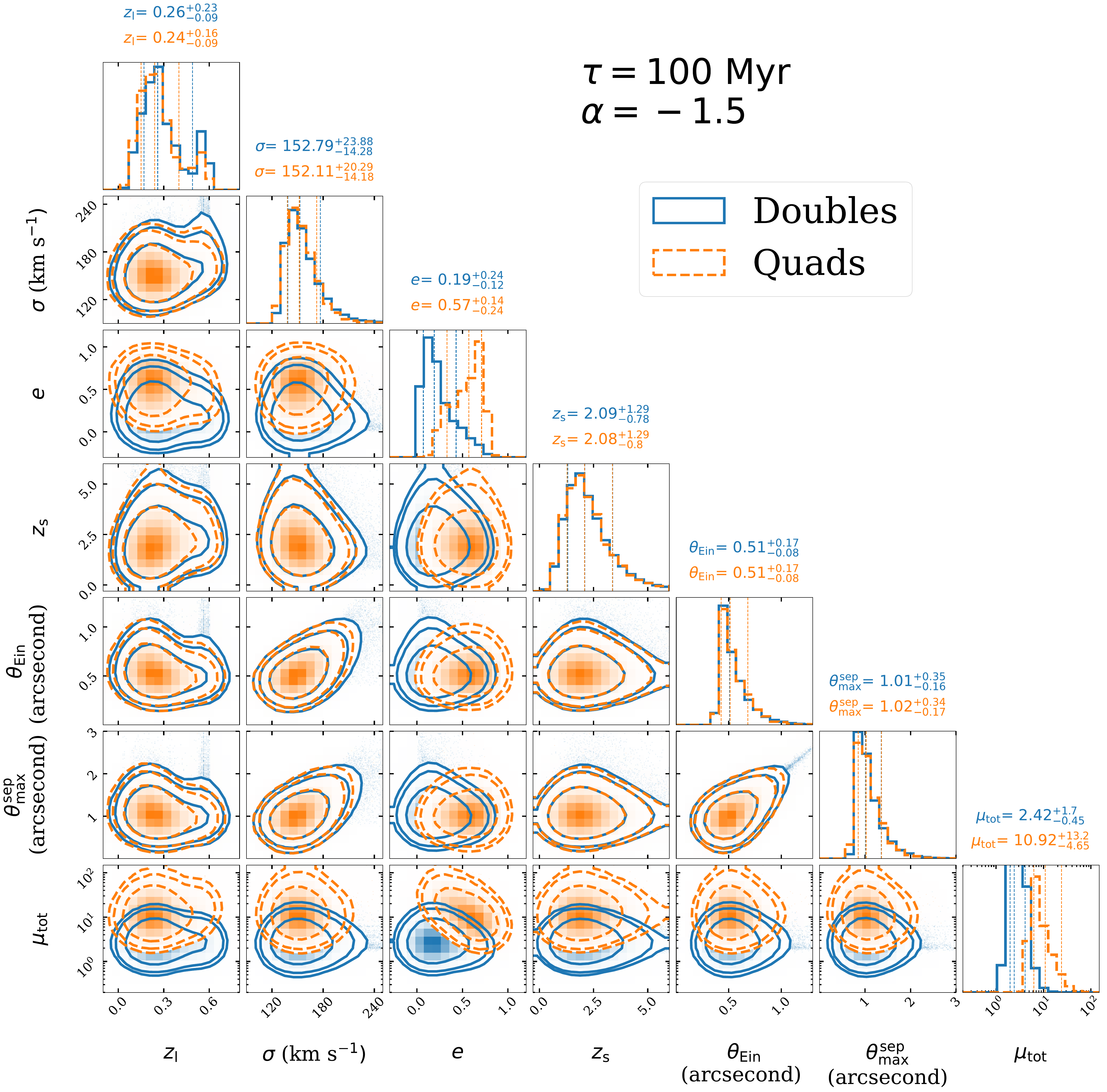}
         {1.0\textwidth}{}}
\caption{Distribution of parameters of double (blue) and quad (orange) lens populations. From left, we show the $\lensred$, lens velocity dispersion ($\sigmavd$), lens ellipticity (e), source redshift ($\srcred$), Einstein radius ($\einrad$), maximum angular separation ($\maxsep$) and magnification ($\mutot$). The contours show the $68 \%$, $90 \%$, and $95 \%$ confidence levels. The dotted vertical lines in the 1D histograms show the median, $16 \%$, and $84 \%$ confidence bounds. As expected, the quads tend to have higher lensing magnifications and arise from more elliptical lenses. } 
\label{fig:dbl_quad_prop}
\end{figure*}
%%%%%%%%%%%%%%%%%%%%%%
%%%%%%%%%%%%%%%%%%%%%%
%%%%%%%%%%%%%%%%%%%%%%%%%%%%%%%%%%%%%%%%%%%%%%%%%%%%%%%%%%%%%%%%%%%%%%%%%%%%%%
For the 10~Myr and 100~Myr scenarios, we find almost 17\%$-$18\% of the initial population is lensed after the simulation. For the 1~Gyr sample, this value decreases to 7\%. Among these lensed KNe, more than 96\% of the population are the doubles. We find that only 3\%$-$4\% of the population leads to quads for each of the three scenarios. 

We note interesting properties of the doubles and quads below. Now, for time-delay cosmography, two of the most important properties are the magnification ($\mu$) of the images and the $\timedelay$ for each of the different images. Hence, we collect all the double and single images from the $\delaytime = 100~\rm{Myr}$ and $\pwrlw = -1.5$ simulation and depict their distribution in the \autoref{fig:dbl_quad_prop}. The blue (orange) distributions show several properties for the doubles (quads) such as - $\lensred$, lens velocity dispersion ($\sigmavd$), $e$, $\srcred$, $\einrad$, maximum angular separation ($\maxsep$) and total magnification ($\mutot$) in different panels. 
We estimate the separation between the different lensed images in the sky-projected plane and define the maximum as $\maxsep$. The total magnification, $\mutot$, is computed as the sum of the $\magnification$ of all lensed images.
Although the double and quad populations exhibit broadly similar distributions for most properties, a notable difference appears in $\mutot$: the quad population contains a larger number of systems with higher $\mutot$ compared to the double population. This trend is consistent with expectations from lensing simulations. These highly magnified quad candidates are of great interest in strongly lensed faint transients. As expected, the lens galaxies with higher ellipticities tend to generate more quads than doubles.
\startlongtable
\tabcolsep3.3pt 
\begin{deluxetable*}{cccccccccc}
\tablecaption{Sumamry of lensed KNe population }
\scriptsize
\tablehead{\colhead{Minimum Delay time }& \colhead{Power-law Slope}
&\colhead{Total lensed Population}
&\colhead{Double}
&\colhead{Quad}
&\multicolumn{5}{c}{Detectable population ($\rm yr^{-1}$)}\\
\colhead{}&\colhead{}
&\colhead{($\rm yr^{-1}$)}&\colhead{($\rm yr^{-1}$)}&\colhead{($\rm yr^{-1}$)}
&\multicolumn{2}{c}{Resolved}
&\multicolumn{2}{c}{Unresolved $^\dagger$}\\
\colhead{($\delaytime$)}&\colhead{($\pwrlw$)}&\colhead{($\%$ of unlensed population)}
&\multicolumn{2}{c}{($\%$ of lensed population)}&
\colhead{30~sec} & \colhead{180~sec}&
\colhead{30~sec} & \colhead{180~sec}\\ 
}
\startdata
10~Myr& $-1.5$&82558(18)&79797(97)&2761(3)&--&5&14&77\\
100~Myr& $-1.5$&65597(17)&63359(97)&2238(3)&--&7&20&87\\
1~Gyr& $-1.5$&13398(7)&12921(96)&477(4)&2&7&7&36\\
\enddata
\tablecomments{
 \scriptsize Rate of lensed KNe, double and quad, and detectable lensed population for three different values of $\delaytime$ and $\alpha$ from \autoref{tab:kn_rate_table}. For all the population, the rate of double images is much higher ($\sim 97 \%$) compared to the quad images. In the resolved and unresolved columns under the detectable population, we mention the rate of lensed KNe, where at least one of the lensed images have peak brightness value higher than 30~sec or 180~sec single-exposure limit in LSST $r$ and $i$ band. $^\dagger$ denotes the images where the $2\einrad<1.5$~arcsec and $\timedelay<3$~days. For these unresolved images, we combined all the $\magnification$ values, as these images will not be resolved in the LSST.
}
\label{tab:lensed_kn_rate_table}
\end{deluxetable*}
%%%%%%%%%%%%%%%%%%%%%%%%%%%%%%%%%%%%%%%%%%%%%%%%%%%%%%%%%%%%%%%%%%%%%%%%%%%%%%
%%%%%%%%%%%%%%%%%%%%%%%%%%%%%%%%%%%%%%%%%%%%%%%%%%%%%%%%%%%%%%%%%%%%%%%%%%%%%%
%%%%%%%%%%%%%%%%%%%%%%%%%%%%%%%%%%%%%%%%%%%%%%%%%%%%%%%%%%%%%%%%%%%%%%%%%%%%%%
%%%%%%%%%%%%%%%%%%%%%%%%%%%%%%%%%%%%%%%%%%%%%%%%%%%%%%%%%%%%%%
\begin{figure*}
\gridline{\fig{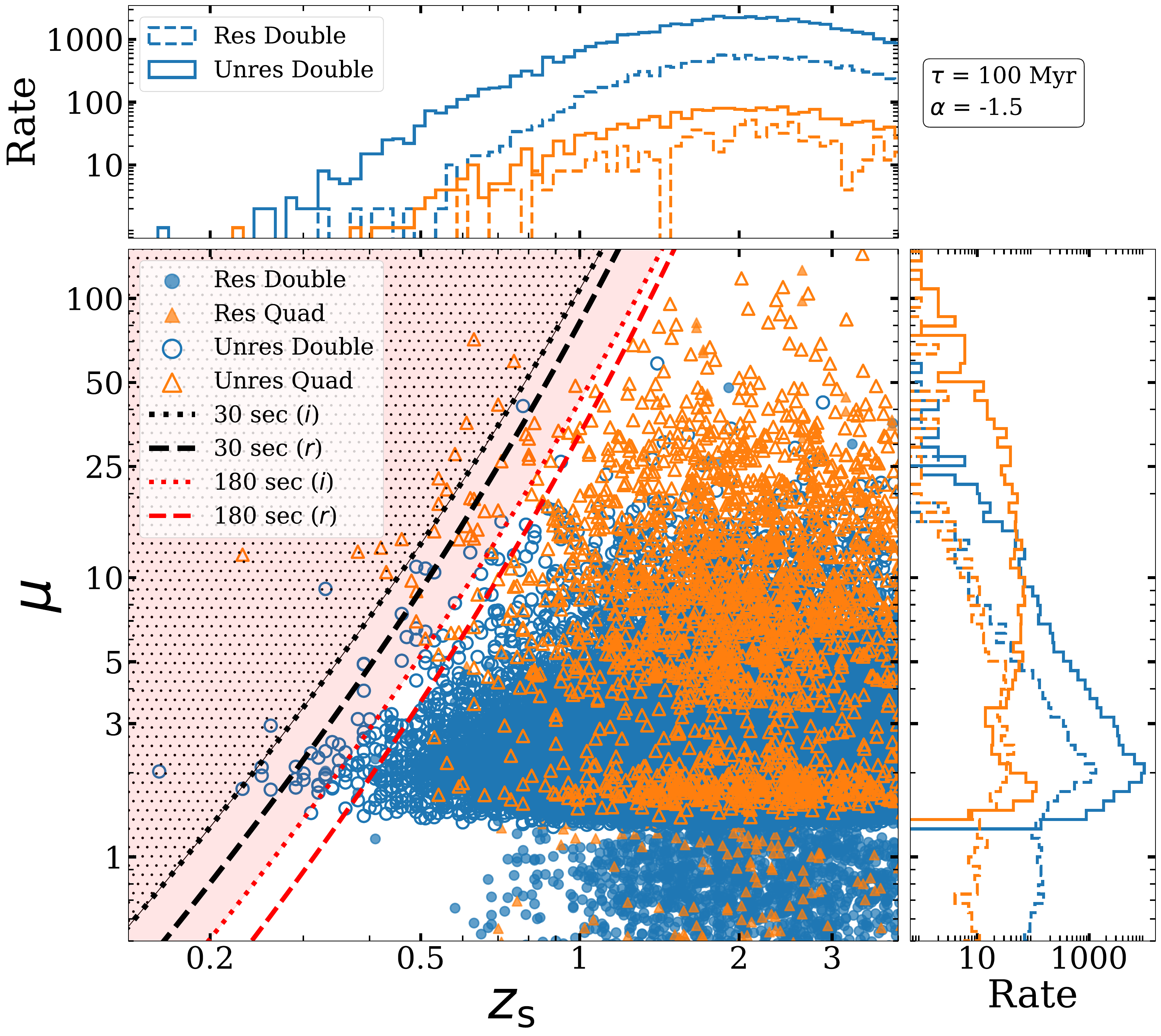}{0.95\textwidth}{}}
\caption{Lensing magnification as a function of the redshift of the lensed KNe. All the blue (orange) colored makers are for the doubles (quads) generated from a DTD with $\delaytime = 100~\rm{Myr}$ and $\pwrlw=-1.5$. The open and filled markers indicate the unresolved and resolved lensed images. For a range of  $\srcred$ values, black (red) curves shows the minimum $\mu$ needed for an AT2017gfo-like lensed KN to be detectable in the LSST $r$ and $i$ bands for a 30~(180)~sec single-exposure limit. Thus, any AT2017gfo-like lensed KN in the dotted (or pink shaded) region will have enough $\mu$ at that $\srcred$ to be detectable.} 
\label{fig:src_lens_redshft_vs_mag}
\end{figure*}
%%%%%%%%%%%%%%%%%%%%%%%%%%%%%%%%%%%%%%%%%%%%%%%%%%%%%%%%%%%%%%
\subsubsection{Detectability of Highly Magnified Lensed KNe}\label{S:highly_magnified}
Here, we explore the features of the highly magnified lensed KNe.
Given the small image separation of most of our lenses (1~arcsec), many of the images will be unresolved and will be detected as a single image with combined magnifications. Therefore, any lens system with $2\einrad<1.5$~arcsec and a maximum $\timedelay<3$~days, provided $\mutot> 1$~\footnote {The condition on $\mutot$ is applied for simplicity.} is assumed to be unresolved. Lenses not satisfying these conditions are considered to be resolved.
In \autoref{fig:src_lens_redshft_vs_mag}, we plot each of the images for the resolved doubles (blue solid circles) and resolved quads (orange solid triangles) in the $\magnification$ vs $\srcred$ plane. Furthermore, for all of the unresolved lenses, we add the $\magnification$ of all of the images to create one single {\it highly} magnified image per lens system (open circles and triangles). In the top and side panels, we show the corresponding 1D distributions of the (un)resolved doubles and quads. As expected, the quads show higher magnifications compared to the doubles, but the fraction of doubles is higher than that of quads. The fraction of unresolved systems is more than that of the resolved ones. 

In order to understand the detectability of any of the lensed population, we need to assume a light curve model. For simplicity, we assume an \afterglow\,-like KN to demarcate the parameter space into detectable vs non-detectable regions.
The dashed and dotted curves show the limiting $\magnification$ needed for an \afterglow\,-like KN to be detectable in the $r$ and $i$ bands, respectively, whereas the colors, black and red, indicate the single-exposure limit of 30~sec and 180~sec, respectively, to determine the detectability. Hence, any images falling in the black dotted (pink shaded) region will be detectable in the $i$ band for the  30~sec (180~sec) single-exposure limit. Most of the detectable lenses comprise {\it unresolved quads} arising from $\srcred\gtrsim0.4$ with moderate-to-high $\magnification\gtrsim10$ and a small fraction of unresolved doubles, albeit from lower redshifts due to their typically low magnifications. For any other KN light curve model, we expect that the dotted (or dashed) near-diagonal curves will shift parallel in the orthogonal direction.      
%%%%%%%%%%%%%%%%%%%%%%%%%%%%%%%%%%%%%%%%%%%%%%%%%%%%%%%%%%%%%%
\begin{figure*}
\gridline{\fig{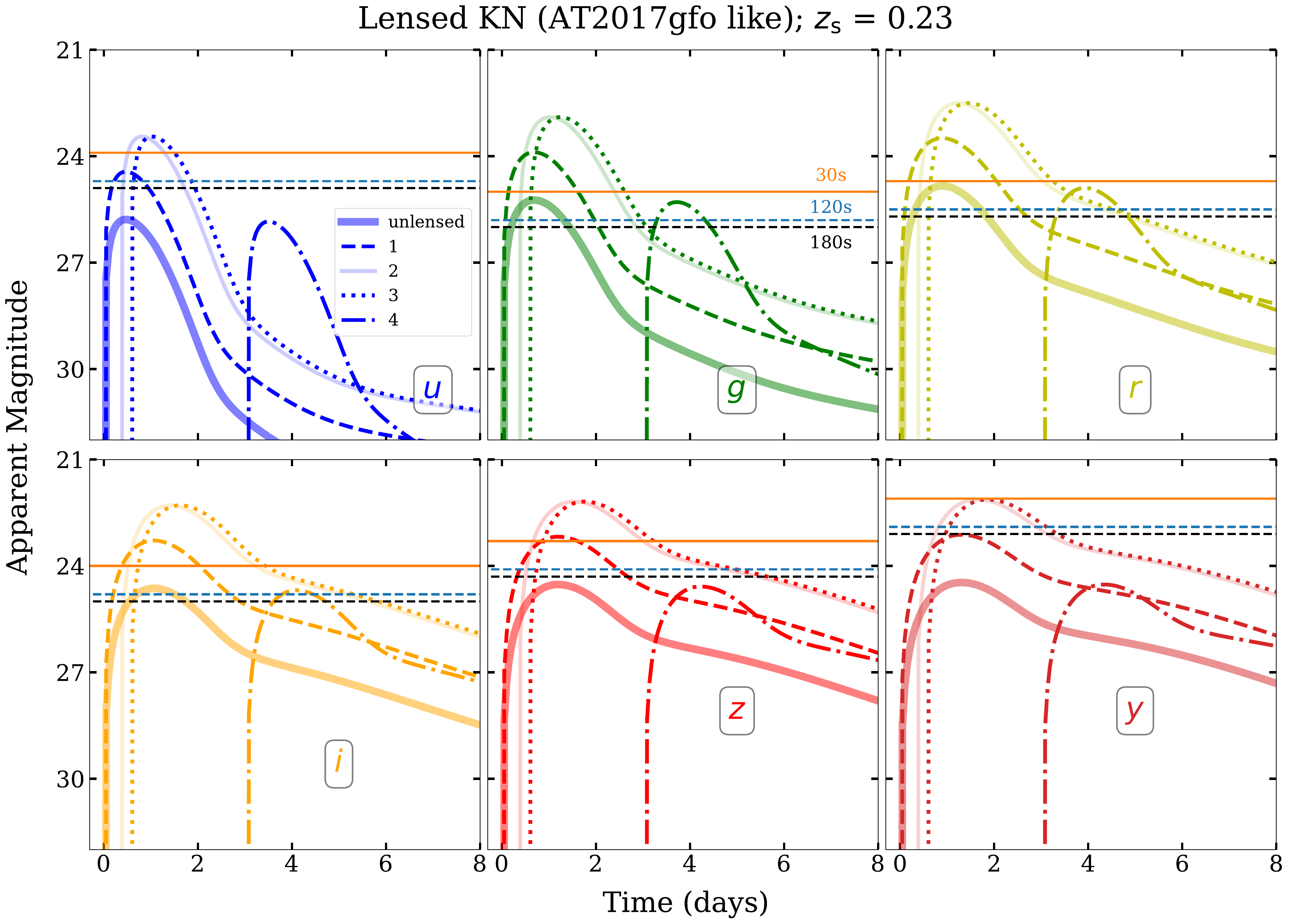}{0.95\textwidth}{}}
\caption{Light curve of a highly magnified lensed KN in all of the six LSST bands. The solid darker curves show the unlensed KN light curve in different bands. Different line styles show light curve of different images after lensing. The three horizontal lines (orange, blue-dashed, black-dashed) show the 30~sec, 120~sec, and 180~sec single-exposure magnitude limits in LSST. After combining  the $\magnification$ values of second and third images (see \autoref{S:highly_magnified}), the peak of the light curve is detectable for all of three single-exposure limits in all of the six bands.} 
\label{fig:high_mag_light_curve}
\end{figure*}
%%%%%%%%%%%%%%%%%%%%%%%%%%%%%%%%%%%%%%%%%%%%%%%%%%%%%%%%%%%%%%

Next, we choose one of the nearest quad lens system ($\srcred=0.23$) to produce the lensed light curves. The second and the third lensed images have $\mu=4.46$ and $\mu=4.15$ with a $\timedelay=5$~hours. We apply a \afterglow\,-like light curve model where the parameter values are taken from \citet{villar_2017_dec}. In \autoref{fig:high_mag_light_curve}, each panel represents one band of LSST. We depict both the unlensed (thick solid curve) and lensed light curves (different line styles). The three  horizontal lines in each panel show the 30~sec (orange), 120~sec (blue dashed), and 180~sec (black dashed) single-exposure limiting magnitude with Rubin. The unlensed light curve is undetectable for a 30~sec exposure limiting for all bands. Assuming that the second and third image are unresolved, the combined magnification leads to a detectable light curve peak in all of the six LSST bands. Fortunately, the fourth image, with a $\timedelay=3$~days, is also detectable in the $g$ and $r$ bands although for the deeper 120~sec and 180~sec single-exposure limits. Early identification of second and third lensed images of such a system can help us to plan follow-up of the fourth image.
%%%%%%%%%%%%%%%%%%%%%%%%%%%%%%%%%%%%%%%%%%%%%%%%%%%%%%%%%%%%%%%
%%%%%%%%%%%%%%%%%%%%%%%%%%%%%%%%%%%%%%%%%%%%%%%%%%%%%%%%%%%%%%%
\subsection{Comparison of lensed KNe with previous studies} \label{S:comparsion}
We compare our work with other studies on lensed KNe from the literature. 
\citet{smith_2023} generated the lensed GW population and evaluated the detectability of their optical counterparts for specific high magnification scenarios in LSST or deeper Rubin follow-up data. While their lensed GW population comprised of doubles only, we are working with statistically realistic proportions of doubles and quads for the lensed KNe population.
The presence of quads can affect the detectability rate of lensed KNe, as quads tend to have higher magnifications. Furthermore, our work does not explore simultaneous GW-KN detection but rather focuses on the chance of detectability of statistically realistic KN samples that will be found in the optical, particularly, in the LSST-WFD survey and possibly in some deeper Rubin data.
In \citet{smith_2023} and our work, lens systems with $\magnification\sim\mathcal{O}(100)$ are likely to yield detectable lensed KNe, although deeper Rubin observations will mostly be required in such cases.

\citet{ma_lensed_kne_2023} also studied the lensed GW population in the context of next-generation GW observatories and found that detection of their electromagnetic counterparts in the LSST-WFD survey will be challenging. As a result, they were advised to look for redder or infrared wavelengths and use the Roman Space Telescope and JWST to discover the electromagnetic counterparts. We also find that detection of lensed KNe in LSST-WFD will be difficult, but for a longer exposure time, such as 120~sec and 180~sec in Rubin, detection of lensed KNe can be possible.

\citet{ma_lensed_kne_2023} also studied the lensed GW population in the context of next-generation GW observatories and found that detection of their electromagnetic counterparts in the LSST-WFD survey will be challenging. As a result, they were advised to look for redder or infrared wavelengths and use the Roman Space Telescope and JWST to discover the electromagnetic counterparts. We also find that detection of lensed KNe in LSST-WFD will be difficult, but for a longer exposure time, such as 120~sec and 180~sec in Rubin, detection of lensed KNe can be possible.
%%%%%%%%%%%%%%%%%%%%%%%%%%%%%%%%%%%%%%%%%%%%%%%%%%%%%%%%%%%%%%%
%%%%%%%%%%%%%%%%%%%%%%%%%%%%%%%%%%%%%%%%%%%%%%%%%%%%%%%%%%%%%%%
%%%%%%%%%%%%%%%%%%%%%%%%%%%%%%%%%%%%%%%%%%%%%%%%%%%%%%%%%%%%%%%
%%%%%%%%%%%%%%%%%%%%%%%%%%%%%%%%%%%%%%%%%%%%%%%%%%%%%%%%%%%%%%%
\section{Summary and Conclusion}\label{S:conclusion}
Our understanding of KNe, their progenitors, and the physics driving the multi-messenger emission is fairly limited. Also, KNe pose a unique challenge in detection and characterisation due to their rarity and faintness compared to the SNe. However, with advanced GW detectors and large optical telescopes such as Rubin, we have the necessary technology to discover GW sources and their optical counterparts, provided sufficient studies are conducted to optimise follow-up strategies and observations to ensure their discoveries. To this end, we simulate realistic populations of both lensed and unlensed KNe to understand their properties and detectability. 

The two main ingredients, while generating our KNe population, are the realistic redshift distribution of the BNS, which affects the KNe rates and the light curve models, which affect their detectability.
We use realistic BNS merger rates from SAF19 to properly incorporate the redshift dependencies in the KNe population. Different binary evolution parameters, corresponding to the DTD, such as $\delaytime$ and $\pwrlw$, help to understand which type of KNe population dominates at particular redshifts. Next, we generate light curves for each of these KNe using \mosfit\, a three-component ejecta model (motivated by the best fit to \afterglow). Using the distribution of ejecta masses, velocity, and opacities, we can include a diverse set of light curves. Thus, we finally create unlensed KNe populations for different DTD scenarios. 

For the lensed KNe, we closely follow the \citet{a_more2022} formalism, which is designed to produce statistically realistic lens populations. We derive the foreground lens properties from the massive early-type galaxies of the HSC survey and use the aforementioned KNe population as our background sources.
However, for simplicity, we generate three scenarios for the lensed populations corresponding to three sets of $\delaytime$ and $\pwrlw$ values (see \autoref{tab:lensed_kn_rate_table}). 
We only analyse those doubles and quads for which the $\einrad>0.4$~arcsec. Below, we describe our key results from the analyses of both the unlensed and lensed KN populations.
\begin{itemize}
    \item We analyse all of the unlensed KNe for all of the scenarios given in \autoref{tab:kn_rate_table}, which are detectable in the LSST $r$ and $i$ bands. Depending on the exposure limits and DTD parameters, we can expect hundreds to thousands of KNe in a year, where the peak magnitude is brighter than the LSST magnitude limits. From \autoref{tab:kn_rate_table}, we notice the ``fixed rate" distribution can produce the largest rate of detectable KNe. The second largest detectable KNe population comes from  $\delaytime = 1~\rm{Gyr}$ and $\pwrlw = -1.5$. The smallest detectable KNe population comes from $\delaytime = 10~\rm{Myr}$ and $\pwrlw = -0.5$. Interestingly, from various $\delaytime$ and $\pwrlw$ values, we notice the farthest detectable KNe arise from $\srcred\sim0.30$. However, after applying realistic LSST observing strategies, some of these rates may decrease. 
    \item The color evolution of the KNe is much faster than the SNe, specifically, the \typeonea\,. We, thus, propose to use colors at $t_{\rm{p}}$, the peak of the light curve, with colors at $t_{\rm{p+3}}$~days as a fast diagnostic tool to distinguish KNe from other SNe at low latencies and can help in triggering suitable follow-up observations. 
    \item For the lensed KNe, we explore three scenarios for the DTD $\delaytime=10~\rm{Myr}, 100~\rm{Myr}, and\, 1~\rm{Gyr}$ and a fixed $\pwrlw=-1.5$. We find the $\delaytime=10~\rm{Myr}$ produces the highest rate of lensed KNe, and $\delaytime=1~\rm{Gyr}$ leads to the lowest rate. However, if we look at the fraction of {it detectable} lensed KNe with respect to the total lensed KNe, population from $\delaytime=1~\rm{Gyr}$ dominated over others.  As expected, the quads show higher $\mutot$ and are in lenses with higher ellipticities compared to the doubles.
    \item To check the detectability of the lensed KNe in different LSST bands, we compared the peak brightness of the magnified light curve with the \afterglow\ event by placing the latter at that particular $\srcred$ of that lensing event. Different curves in \autoref{fig:src_lens_redshft_vs_mag} set up the limiting value of $\mu$ needed for an \afterglow\,-like event to have peak brightness higher than different single-exposure limits in the LSST $r$ and $i$ bands for a range of $\srcred$ values. For example, in the case of $\delaytime = 100~\rm{Myr}$, we notice 9 (64) of such lens systems have peak brightness value higher than the 30 (180)~second single-exposure limit of both $r$ and $i$ band.
    \item For some lens systems, where the 2$\einrad$ is less than 1.5~arcsec and also the $\timedelay$ of different images is less than the 3~ days, we combine the $\mu$ of all of those images. These combined and magnified (green triangles in \autoref{fig:src_lens_redshft_vs_mag}) images become particularly interesting due to the increase of the peak brightness, which brings them into the detectable range of LSST $r$ and $i$ bands. For all three different BNS merger rate parameters in \autoref{tab:lensed_kn_rate_table}, there are higher rates of detectable cases in the combined population, in comparison to the non-combined cases (see green triangles in \autoref{fig:src_lens_redshft_vs_mag}).  
\end{itemize}

We will explore broader implications of this kind of analysis in different types of transients with the real LSST data. With the availability of the recent LSST DP1 data, we are already extracting potentially massive elliptical lensing deflector galaxies from different observed fields to replace the HSC galaxies in our lensing simulations. Moreover, we will also make real image cutouts for these types of lens systems by injecting images of lensed transients around the LSST galaxies. These types of image cutouts are particularly useful for different types of image analysis algorithms. In future, we also plan to explore the host galaxy properties and their effects on the lensed KNe source population. Different host galaxy properties, such as age, metallicity, extinction, and offset in KNe position due to natal kicks, can further help to constrain the BNS merger rate, KNe light curve, and also the parameters of lensing simulations. It is also important to mention that the microlensing effect due to the stellar population present in the lensing galaxies can affect the $\mu$ of the lensed systems and hence can change the final light curve magnitude. This type of microlensing effect is particularly interesting in the field of strongly lensed SNe and has gained focus in recent years \citep{dobler_keeton2006, goldstein2018,huber2021}. For example, in the context of LSST, \citet{arendse2024} predicted that microlensing effects can decrease the detection of lensed SNe events by 8 $\%$. Due to current limitations, we were unable to include the microlensing effect in our current lensed KNe simulations.

In this work, we have investigated the detection and identification of (un)lensed KNe in LSST, while creating a population of KNe from realistic BNS merger rate distributions and light curve parameters. In this study, we find that different binary evolution parameters can affect the final rates of detectable (un)lensed KNe population in LSST. We show that the color evolution of light curves and color-color diagram can help us in the detection and classification of KNe events from other types of transients. In the case of lensed KNe, we observe that a handful of \afterglow-like events can have double and quad images, where the $\mu$ value after lensing can bring them in the observable regime of different LSST bands.
\section*{Data Availability}
The data are available from the authors upon reasonable request after the publication of the work.
\section*{acknowledgments}
We thank Prajakta Mane and Surhud More for scientific discussions and useful advice.
We acknowledge the use of the high-performance computing facility Pegasus at IUCAA to carry out all the simulations. We acknowledge the support from the SERB Power Grant (SPG/2022/001866) funded by DST.

\textit{Software} : For this work, we extensively use many open software projects like Python 3 \citep{python3}, Numpy \citep{numpy_scipy_matp}, scipy \citep{numpy_scipy_matp}, matplotlib \citep{numpy_scipy_matp}, pandas \citep{numpy_scipy_matp} and Astropy \citep{astropy:2013, astropy:2018}
%%%%%%%%%%%%%%%%%%%%%%%%%%%%%%%%%%%%%%%%%%%%%%%%%%%%%%%%%%%%%%%
%%%%%%%%%%%%%%%%%%%%%%%%%%%%%%%%%%%%%%%%%%%%%%%%%%%%%%%%%%%%%%%
\appendix
%%%%%%%%%%%%%%%%%%%%%%%%%%%%%%%%%%%%%%%%%%%%%%%%%%%%%%%%%%%%%%%
%%%%%%%%%%%%%%%%%%%%%%%%%%%%%%%%%%%%%%%%%%%%%%%%%%%%%%%%%%%%%%%
\section{Kilonova Physical Properties}\label{S:app_kne_physical_properties}
\begin{figure*}
\gridline{\fig{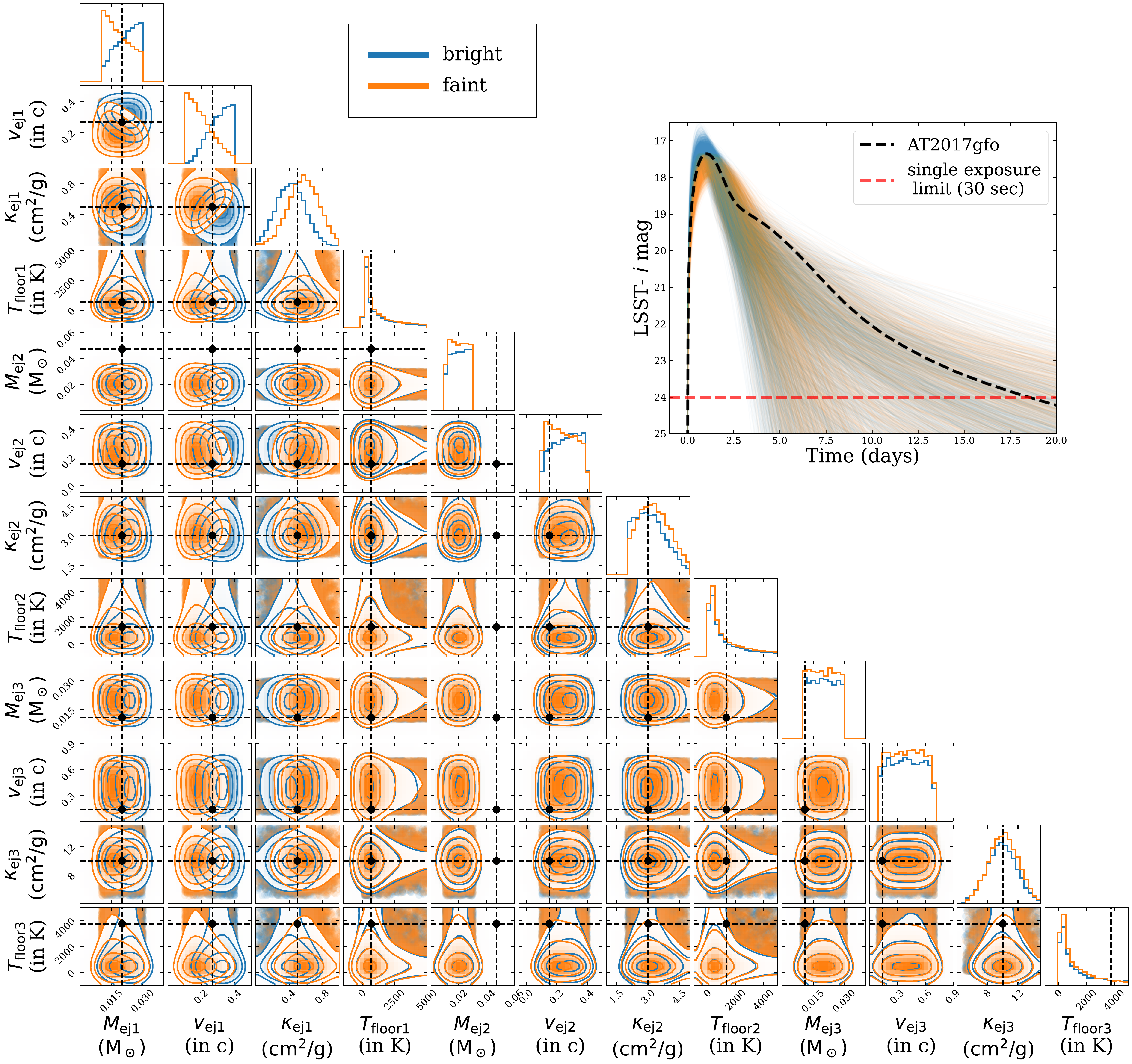}{1.0\textwidth}{}}
\caption{Correlation and 1-d distribution of ejecta properties of KNe. The blue (orange) color denotes the bright (faint) population compared to the \afterglow. The black dashed line and dot represent the light curve and the physical parameters of the \afterglow. Horizontal red dashed line shows the 30~sec single-exposure limit for the LSST $i$ band.} 
\label{fig:kn_prop_corner_plt}
\end{figure*}

From \autoref{tab:all_kn_para_table}, we have generated a sample of KNe for different physical properties at the $\srcred$ of \afterglow. In this sample, if a particular KN has a higher peak brightness compared to \afterglow\, we denote it as a bright KN. \footnote{This bright population is not related to our bright and detectable KNe population in the LSST. This division is made only to compare with \afterglow.} In \autoref{fig:kn_prop_corner_plt}, we show the correlation and 1-d histograms of physical parameters for the bright (faint) KNe population in blue (orange) color.   The dashed black line and the dot correspond to the physical parameters for the \afterglow\,. The red dashed line shows the 30~sec single-exposure limit of the LSST $i$ band. We notice that for KNe, which have a peak brightness higher than \afterglow\, they also tend to have a higher value of $\mejone$ and $\vejone$. The $\kappaone$ value is also low for the bright population compared to the faint group. $\mejtwo$ value is particularly high in the case of \afterglow\, which results in a small increase in the brightness in the middle region (around 3~days to 6~days) of the light curve. We do not enforce this high value of $\mejtwo$ in our KNe light curve model. In conclusion, we observe that the peak brightness of the KNe light curve is mainly controlled by the blue ejecta mass ($\mejone$) and velocity ($\vejone$).
%%%%%%%%%%%%%%%%%%%%%%%%%%%%%%%%%%%%%%%%%%%%%%%%%%%%%%%%%%%%%%%
%%%%%%%%%%%%%%%%%%%%%%%%%%%%%%%%%%%%%%%%%%%%%%%%%%%%%%%%%%%%%%%
\section{Lensed Kilonovae for varying delay time}\label{S:app_all_dbls_qds_pop}
\begin{figure*}
\gridline{\fig{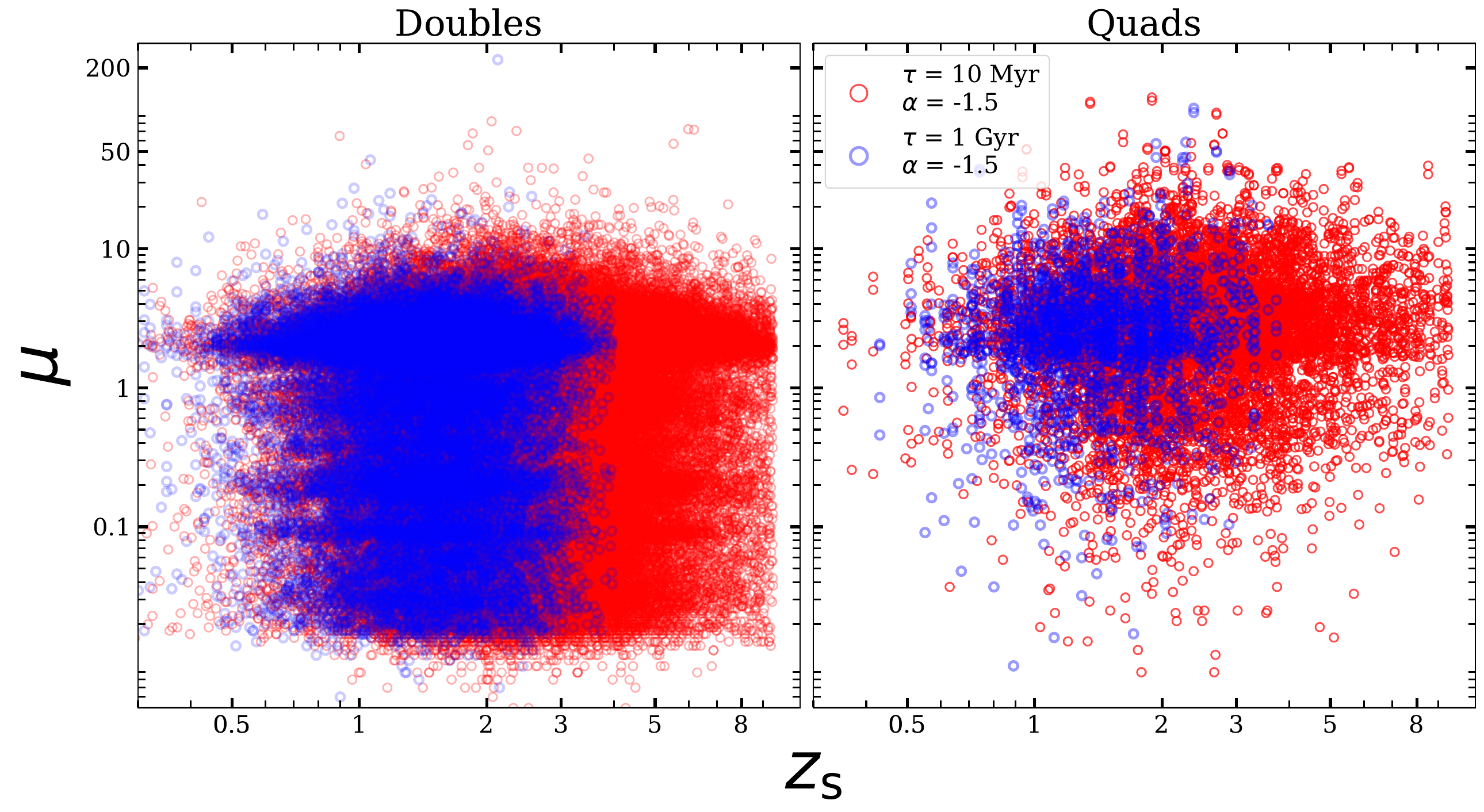}{1.0\textwidth}{}}
\caption{ Distribution of $\magnification$ of double and quad lensed images with $\srcred$ for two different $\delaytime$ values. Blue (red) colored circles denote the $\delaytime =  1 (10)\rm{~Gyr~(Myr)}$ in both panels. 
Long $\delaytime$ makes the BNS evolution and merger timescales longer.} 
\label{fig:app_two_lensed_pop}
\end{figure*}

We explored the impact of varied delay time of the DTD for a fixed $\pwrlw=-1.5$. 
In \autoref{fig:app_two_lensed_pop}, we show how the $\magnification$ of the doubles and quads changes with $\srcred$ as a result. The red (blue) open circles are the lensed KNe for the delay time parameter of DTD of the BNS merger rate, $\delaytime = 10~\rm{Myr}~(1~\rm{Gyr})$. The quads in both cases have higher magnification compared to the doubles. The lensed KNe population also peaks at different $\srcred$ for different $\delaytime$, similar to the unlensed KNe. Long $\delaytime$ creates a lensed KNe population, which peaks at lower $\srcred$ compared to shorter $\delaytime$. This can directly favour the detection of lensed KNe in the LSST that comes from the $\delaytime = 1~\rm{Gyr}$ over the $\delaytime = 10~\rm{Myr}$   population.     
%%%%%%%%%%%%%%%%%%%%%%%%%%%%%%%%%%%%%%%%%%%%%%%%%%%%%%%%%%%%%%%
%%%%%%%%%%%%%%%%%%%%%%%%%%%%%%%%%%%%%%%%%%%%%%%%%%%%%%%%%%%%%%%
\bibliography{main}{}
\bibliographystyle{aasjournal}
%%%%%%%%%%%%%%%%%%%%%%%%%%%%%%%%%%%%%%%%%%%%%%%%%%%%%%%%%%%%%%%
\end{document}